\documentclass[12pt,preprint]{emulateapj}
\usepackage{natbib}

\newcommand{\be}{\begin{equation}}
\newcommand{\ee}{\end{equation}}
\newcommand{\ba}{\begin{eqnarray}}
\newcommand{\ea}{\end{eqnarray}}

\begin{document}

\title{The Near-IR Background Intensity and Anisotropies During The Epoch of Reionization}

\author{Asantha Cooray$^{1}$, Yan Gong$^1$, Joseph Smidt$^1$, Mario G. Santos$^2$}

\affil{$^1$Department of Physics \& Astronomy, University of California, Irvine, CA 92697}
\affil{$^2$CENTRA, Instituto Superior T\'ecnico, Technical University of Lisbon, Lisboa 1049-001, Portugal}

\begin{abstract}
A fraction of the extragalactic near-infrared (near-IR) background light involves redshifted photons from the ultraviolet (UV) emission from galaxies present during
reionization at redshifts above 6.  The absolute intensity and the anisotropies of the near-IR background
provide an observational probe of the first-light galaxies and their spatial distribution.
We estimate the extragalactic background light intensity during reionization by accounting for the stellar and nebular emission from first-light galaxies.
We require the UV photon density from these galaxies to generate a reionization history that is
consistent with the optical depth to electron scattering from cosmic microwave background measurements.
We also require the bright-end luminosity function of galaxies in our models to reproduce the measured Lyman drop-out 
luminosity functions at redshifts of 6 to 8. The absolute intensity is about 0.1 to 0.4 nW m$^{-2}$ sr$^{-1}$ at the peak of its spectrum at $\sim$ 1.1 $\mu$m.
We also discuss the anisotropy power spectrum of the near-IR background using a halo model to describe the galaxy distribution.
We compare our predictions for the anisotropy power spectrum to existing measurements from deep near-IR imaging data from
{\it Spitzer}/IRAC, {\it Hubble}/NICMOS, and {\it AKARI}.  The predicted rms fluctuations at tens of arcminute angular scales are roughly an order of magnitude
smaller than the existing measurements. While strong arguments have been made that the measured fluctuations do not have an origin involving faint
low-redshift galaxies, we find that measurements in the literature are also incompatible with galaxies present during the era of reionization. The 
measured near-IR background anisotropies remain unexplained with an unknown origin.
\end{abstract}

\maketitle

\section{Introduction}

The optical and UV radiation from sources present during reionization is expected to leave a signature in the
extragalactic background light (EBL) at near-IR wavelengths (e.g., Santos et al. 2002; Salvaterra \& Ferrara 2003; Cooray \& Yoshida 2004; Fernandez \& Komatsu 2006; Raue 2009). Such radiation is not expected to be present in the
background light at UV and optical wavelengths due to the redshifted Lyman limit. The exact intensity from first-light galaxies present during reionization is currently unknown. The first predictions 
suggested an intensity as high as 10 to 30 nW m$^{-2}$ sr$^{-1}$ (Santos et al. 2002; Salvaterra \& Ferrara 2003). These estimates were  partly motivated by the need to explain the difference between DIRBE 
EBL measurements (e.g., Cambr\'esy et al. 2001) and  the integrated galaxy light (IGL) from deep galaxy counts  (Madau  \& Pozzetti 2000; Totani et al. 2001). 

These predictions with high backgrounds were questioned by Madau \& Silk (2005) based on existing limits related to metal content at high redshifts and the X-ray background produced by stellar end-products such as black holes.
They suggest an intensity less than about 2.5 nW m$^{-2}$ sr$^{-1}$ in the J-band from a galaxy population made up of population III stars during reionization (Madau \& Silk 2005).
With a combination of Population II stars and changes to the lifetime of stars, Fernandez \& Komatsu (2006) argued that the background could be as high as 4 to 8 nW m$^{-2}$ sr$^{-1}$. 
Even in such a scenario a simple estimate of the UV photon density at $z >6$ shows that there are roughly an order of magnitude higher number of H-ionizing photons per baryon during reionization than necessary to explain the
reionization history. Since one does not expect more than a few H-ionizing photons per baryon during reionization, a first-order estimate suggests that 
the background intensity cannot be larger than a few tenths  nW m$^{-2}$ sr$^{-1}$ between 1 and 2 $\mu$m.

Unfortunately a direct search for the integrated intensity of galaxies  present during reionization
based on absolute background measurements has been problematic due to the confusion with the Zodiacal foreground. At 1 AU Zodiacal light is two to three orders
of magnitude brighter than the $\sim$ 10 nW m$^{-2}$ sr$^{-1}$ intensity produced by extragalactic sources.  While challenging, techniques have been devised to estimate the Zodiacal dust column density 
based on the line strengths of Fraunhofer lines seen in the dust-scattered Solar spectrum (e.g., Bernstein \& Dyson 2003). Instead of the absolute background, in Cooray et al. (2004; also Kashlinsky et al. 2004),  it was
proposed that the galaxies present during reionization can be studied with anisotropies of the near-IR background. The anisotropy studies have the potential to
probe deeper than the absolute experiments and could study a galaxy population present during reionization that leads to an intensity well below 0.1 nW m$^{-2}$ sr$^{-1}$ (Cooray et al. 2004).

This suggestion has motivated experimental measurements on the near-IR anisotropy power spectrum with data from {\it Spitzer}/IRAC, {\it HST}/NICMOS, and {\it AKARI}.
After a deep removal of point sources, Kashlinsky et al. (2005, 2007, 2012) claimed a detection of first-light galaxy fluctuations at $z > 8$. The detected signal is an excess of clustering power above shot-noise on the largest angular scales.
A similar suggestion was also made by an {\it AKARI} group (Matsumoto et al. 2011), but an analysis of the {\it HST}/NICMOS Ultra  Deep Field led to an opposite conclusion that the
sources contributing to the near-IR excess fluctuations are at $z < 8$ (Thompson et al. 2007).
Due to the limited areas of existing deep surveys near-IR background anisotropy measurements are limited to angular scales less than about one degree. The limited field of view is especially a problem for existing
NICMOS UDF measurements (Thompson et al. 2007), where the fluctuations are limited to angular scales less than 5$'$. Separately, a joint analysis of IRAC and {\it HST}/ACS data in the same GOODS fields as studied by Kashlinsky et al. (2007)
led to the suggestion that up to 50\% of the excess fluctuations at 3.6 $\mu$m could come from faint dwarf galaxies at $z \sim 2$ (Cooray et al. 2007; Chary et al. 2008). 
Through detailed models combined with more recent measurements of faint galaxy clustering, Helgason et al. (2012) has lowered this low-redshift contribution to 3.6 $\mu$m intensity fluctuations to be at most  20\%. 
The rest of the anisotropies continue to be interpreted as originating from first-light galaxies during reionization (Kashlinsky et al. 2012).

While there are still uncertainties on the exact intensity and the amplitude of intensity fluctuations in experimental measurements, the situation is no different on the theory side.
The first estimates on the anisotropy power spectrum made use of linear theory clustering (Cooray et al. 2004).
Fernandez et al. (2010) used numerical simulations of reionization to predict the expected power spectrum during reionization. Their power spectrum has a shape
in the form of a power-law with $C_l \propto l^{1/2}$ between ten arcminute to arcsecond angular scales. 
Fernandez et al. (2010) suggested that the power-law behavior arises from significant non-linear biasing of dark matter halos at high redshifts. Due to the limited box sizes of existing  reionization simulations of the order 100 to 140 Mpc
on the side, numerical studies are limited to angular scales of 30 arcminutes and below at $z > 6$.

With the availability of WFC3 on the {\it Hubble} Space Telescope, dedicated IR background experiments (e.g., CIBER sounding-rocket experiment; Bock et al. 2006),
and plans for a future space-based absolute intensity measurement (ZEBRA\footnote{http://zebra.caltech.edu}; Cooray et al. 2009) there is now
a clear need to revisit theoretical predictions on  both the absolute intensity and the anisotropy power spectrum from galaxies present during reionization. While current multi-band {\it Spitzer} and {\it AKARI} 
measurements do not overlap in the same fields, the combination of IRAC and WFC3 data on some of the same well-studied fields on the sky (e.g., fields covered by the CANDELS survey; Grogin et al. 2011; Koekemoer et al. 2011)
will soon allow the spectral energy distribution (SED) of intensity fluctuations be studied uniformly.
Separately CIBER is conducting spectral imaging absolute measurements between 0.8 and 1.6 $\mu$m in wide 4 deg.$^2$ fields instantaneously using multiple sounding rocket flights (Zemcov et al. 2012).
The combination of IRAC and CIBER is capable of extending anisotropy measurements out to angular scales of more than a degree from optical to 4.5 $\mu$m.

In this work we establish both the mean intensity and the anisotropy power spectrum of galaxies present during reionization. We update Cooray et al. (2004) by taking into account
recent developments in the study of reionization, and by introducing a halo model to calculate the non-linear clustering of the IR background intensity. The stellar and nebular emission from first light galaxies follow the calculations
presented in Fernandez \& Komatsu (2006), but we specifically require that the UV photon background produced by the 
galaxy population present from $z \sim 6 $ to 30 is consistent with the optical depth to electron scattering as measured by the WMAP polarization data with a value of 0.088 $\pm 0.014$ (Komatsu et al. 2011).
We account for the current uncertainty in the optical depth by introducing variations to the fiducial model so that the optical depth to electron scattering varies between 0.07 and 0.1.
Our models are also designed to reproduce the bright-end galaxy luminosity functions (LFs) in deep HST/WFC3 surveys at $z > 6$ involving the Lyman-dropout galaxy samples. This normalization at the bright-end of galaxy luminosities
puts strong constraints on the intensity.

This paper is organized as follows. In Section~2 we outline our model for the reionization galaxies including stellar nebulae and the IGM emission. Section~3 presents the calculation
related to luminosity mass density of these galaxies. In Section~4, 5 and 6 we outline the background intensity and spatial anisotropy power spectrum calculations, respectively.
In Section~7 we discuss our results related to the intensity and angular power spectrum and present a comparison to existing measurements. We conclude with a summary in Section~8.
We assume the flat $\Lambda$CDM model with $\Omega_{\rm m}=0.27$, $\Omega_{\rm b}=0.046$, $\Omega_{\Lambda}=0.73$, $h=0.71$, $\sigma_8=0.81$ and $n_{\rm s}=0.96$ \citep{Komatsu11} throughout the paper.

\section{Emission from stars and the intergalactic medium}

We first describe the emission from stars in first-light galaxies present during reionization.
Following Fernandez \& Komatsu (2006), we consider two stellar populations in this calculation. 
The first, referred to as Pop II stars, are metal-poor stars with metallicity $Z=1/50Z_{\odot}$, 
and the second, Pop III stars, are metal-free stars with $Z=0$. 

To describe the stellar initial mass function  (IMF) we make use of two descriptions.
For Pop II stars, we adopt the IMF given by Salpeter (1995)
\be
f(M_*) \propto M_*^{-2.35},
\ee
with mass range from 3 to 150 $M_{\sun}$. For Pop III stars we use the IMF obtained by
Larson (1999), which takes the form as
\be
f(M_*) = M_*^{-1}\left( 1+\frac{M_*}{M_*^{c}} \right)^{-1.35},
\ee
where $M_*^{c}=250\ M_{\sun}$, and the mass range is from 3 to 500 $M_{\sun}$. 

we utilize the fitting results from \cite{Lejeune01} and \cite{Schaerer02} to calculate 
other stellar parameters, such as the intrinsic bolometric luminosity $L_*^{\rm bol}(M_*)$,
the effective temperature $T_*^{\rm eff}(M_*)$, the main-sequence lifetime $\tau_*$, 
and the time-averaged hydrogen photoionization rate $\overline{Q}_{\rm HI}(M_*)$. 
The fitting forms of these parameters are different for Pop II and Pop III stars. 
For Pop II stars, they are given as
\ba
{\rm log_{10}}(L_*^{\rm bol}/L_{\odot}) &=& 0.138+4.28x-0.653x^2, \nonumber \\
{\rm log_{10}}(T_*^{\rm eff}/{\rm K}) &=& 3.92+0.704x-0.138x^2,  \nonumber \\ 
{\rm log_{10}}(\tau_*/{\rm yr}) &=& 9.59-2.79x+0.63x^2, \nonumber \\ 
{\rm log_{10}}(\overline{Q}_{\rm HI}/s^{-1}) &=& 27.80+30.68x-14.80x^2+2.50x^3,\nonumber
\ea
where $x={\rm log_{10}}(M_*/M_{\odot})$ and for Pop III stars, they are
\ba
{\rm log_{10}}(L_*^{\rm bol}/L_{\odot})  &=& 0.4568+3.897x-0.5297x^2, \nonumber \\
{\rm log_{10}}(T_*^{\rm eff}/{\rm K}) &=& 3.639+1.501x-0.5561x^2+0.07005x^3, \nonumber \\
{\rm log_{10}}(\tau_*/{\rm yr}) &=& 9.785-3.759x+1.413x^2-0.186x^3, \nonumber 
\ea
\ba
{\rm log_{10}}(\overline{Q}_{\rm HI}/s^{-1}) = \left\{\begin{array}{ll}
39.29+8.55x\ & \textrm{5-9}\ \rm M_{\odot} \nonumber \\
43.61+4.90x-0.83x^2\ & \textrm{9-500}\ \rm M_{\odot} \end{array}\ .\right.
\ea

From these expressions the stellar radius $R_*(M_*)$ is
\be
4\pi R_*^2(M_*) = \frac{L_*^{\rm bol}(M_*)}{\sigma T_*^{\rm eff}(M_*)^4},
\ee
where $\sigma = 5.67 \times 10^{-5} {\rm \ erg \ s}^{-1} {\rm \ cm}^{-2} {\rm \ K}^{-4}$ is 
the Stefan-Boltzmann constant.  The stellar radius is useful for the calculation related to the stellar emission spectrum (see Section~3.1).

The ionization volume in the nebulae surrounding the stars  (Str$\rm \ddot{o}$mgren sphere) 
can be derived if assuming ionization equilibrium where the ionization rate equals recombination rate
\be \label{eq:Vneb}
V_{\rm ion}^{\rm neb} = \frac{\overline{Q}_{\rm HI}(M_*)}{n_e^{\rm neb}n_{\rm HII}^{\rm neb}\alpha_{\rm B}^{\rm rec}},
\ee
where $\alpha_{\rm B}^{\rm rec}$ is the hydrogen case B recombination coefficient
which depends on the gas temperature $T_{\rm gas}$ (assumed to be $\simeq 3\times 10^4$ K), 
and we will discuss it in detail in the next Section. 
Here, $n_e^{\rm neb}$ and $n_{\rm HII}^{\rm neb}$ are the local number density of electron and 
HII in the stellar nebulae where we assume $n_e^{\rm neb}=n_{\rm HII}^{\rm neb}=10^4\ \rm cm^{-3}$.

For the IGM, the hydrogen density is lower than that of the stellar nebulae, so we no longer
assume the ionization equilibrium. We estimate the ionization volume by a redshift-dependent 
form as \citep{Santos02}
\be \label{eq:VIGM}
V_{\rm ion}^{\rm IGM}(z) = \frac{\overline{Q}_{\rm HI}(M_*)}{\bar{n}_{\rm H}(z)}\tau_*,
\ee
where $\bar{n}_{\rm H}(z)=1.905\times 10^{-7}(1+z)^3\ \rm cm^{-3}$ is the mean
hydrogen number density for $\Omega_b=0.046$ assumed in this work (Shull et al. 2011).

These quantities discussed here would now be used  to estimate the luminosity mass
density, near-IR background intensity SED, and the anisotropy power spectrum.  We make use of emission from
 the stellar nebulae and the IGM for both Pop II and Pop III stars in  galaxies present during reionization.

\section{Luminosity mass density of the sources}

\begin{figure*}[htb]
\centerline{
\includegraphics[scale = 0.4]{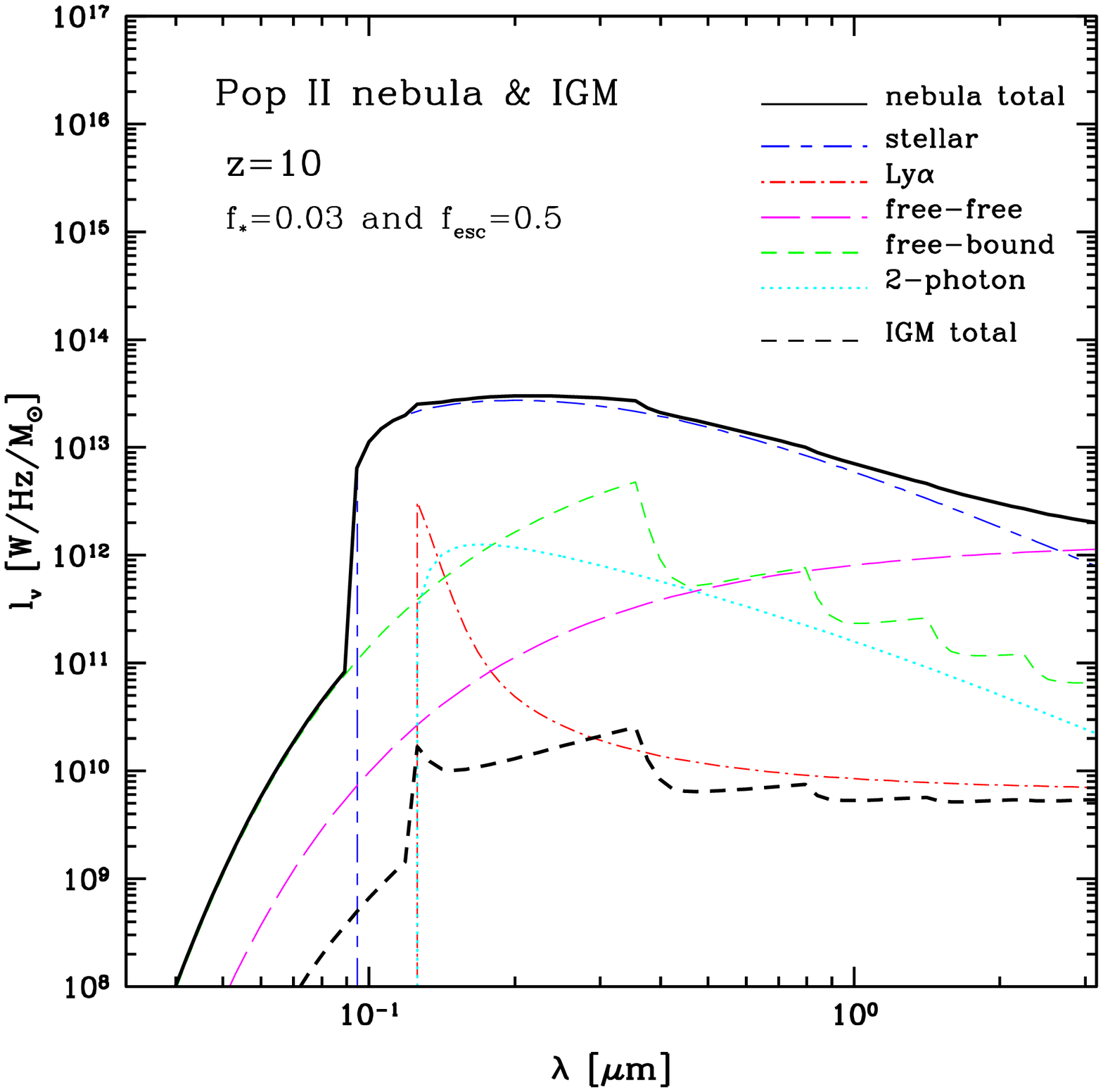}
\includegraphics[scale = 0.4]{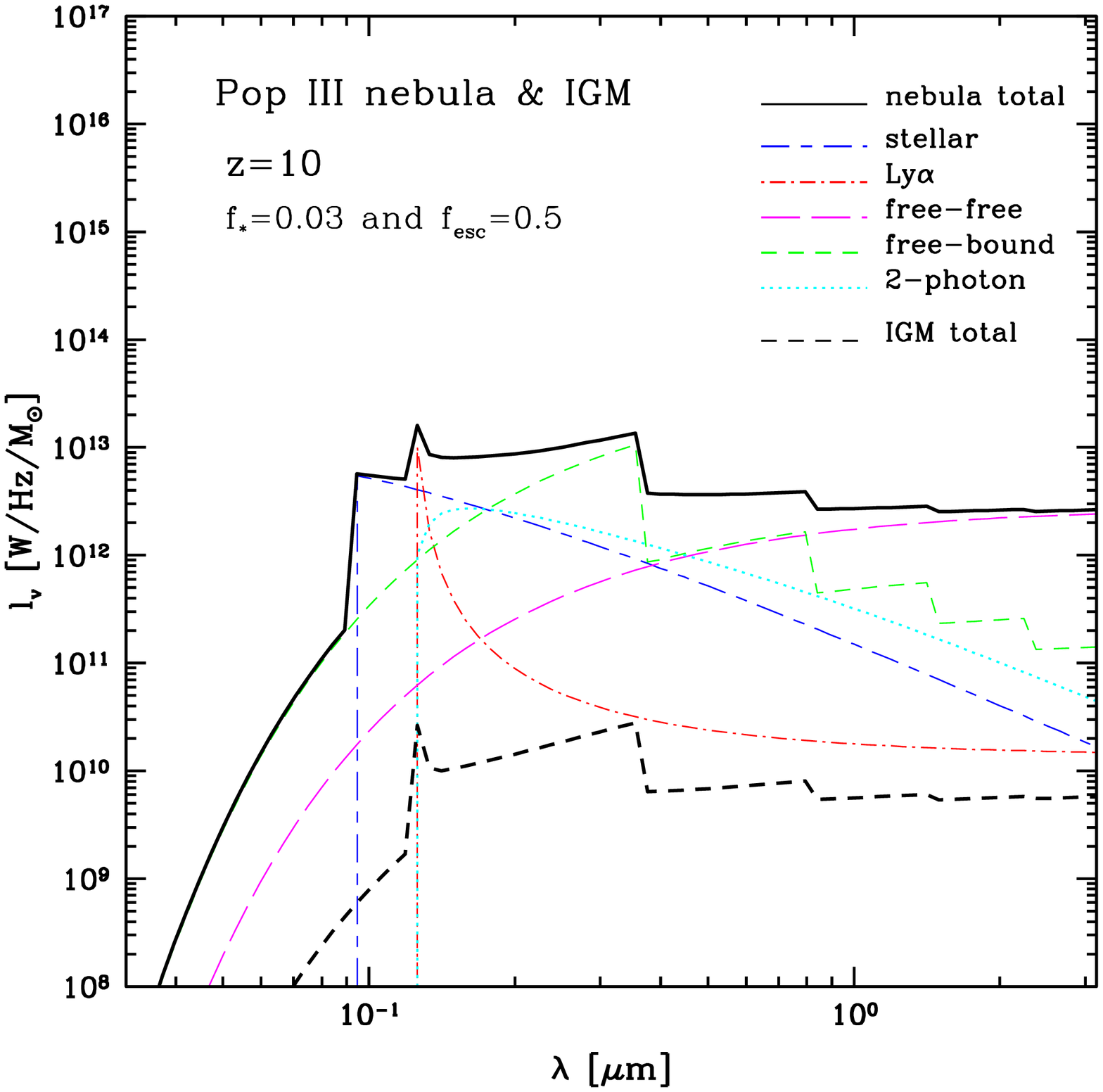}
}
\caption{\label{fig:Lv} The luminosity mass density $l_{\nu}$ vs. the rest-frame wavelength 
$\lambda$ for the Pop II and Pop III stars at $z=10$. The total $l_{\nu}$ from the stellar nebula and the IGM 
are shown respectively in each figure (thick lines). Also, to specify the contribution from different sources, 
we plot the $l_{\nu}$ of stellar, Lyman-$\alpha$, free-free, free-bound and two-photon emission for the stellar 
nebula case (thin lines). We take $f_*=0.03$ and $f_{\rm esc}=0.5$ for both Pop II and Pop III cases. 
We find the $l_{\nu}$ of the stellar nebula is much larger than that of the IGM for both Pop II 
and Pop III stars.}
\end{figure*}

In this Section, we calculate the luminosity per stellar mass at frequency $\nu$, i.e. luminosity mass density,
for several sources that contribute to the infrared background, such as the direct emission from the stars,
Lyman-$\alpha$ line, and free-free, free-bound and two photon processes. The luminosity mass density takes the central role in our
estimation of the near-IR background intensity spectrum.

\subsection{Stellar spectrum}

For simplicity, we assume the stellar spectrum is a Planckian truncated
at $h\nu=13.6$ eV. Thus, the stellar luminosity at frequency $\nu$ can 
be expressed as
\ba
L_{\nu}^* = \left\{\begin{array}{ll}
\pi S_*B_{\nu}(T_*^{\rm eff})\ & \textrm{for}\ h\nu<13.6\ \rm eV\\ \nonumber
0\ & \textrm{for}\ h\nu\ge13.6\ \rm eV \end{array}\ ,\right.
\ea
where $S_*=4\pi R_*^2$ is the surface area of the star, $R_*$ is the stellar radius, $M_*$ is the stellar mass and $B_{\nu}(T)$ 
is the Planck spectrum
\be
B_{\nu}(T_*^{\rm eff})=\frac{2h\nu^3/c^2}{e^{h\nu/kT}-1}.
\ee
Note that for simplicity we have ignored the absorption lines of the Lyman-alpha (Ly$\alpha$) series 
here. The absorption is not strong enough to affect the shape of the spectrum
and hence it is not expected to affect our results especially in the infrared wavelengths \citep{Santos02,Fernandez06}. Our predictions related to the optical
background fluctuations may be somewhat overestimated. Also, the emission with $h\nu\ge13.6$ eV cannot be approximated by a black-body spectrum, 
thus we use the fitting formulae for time-averaged photoionization rate, $\overline{Q}_{\rm HI}$, for
Pop II and Pop III stars to calculate the emission at higher energies.

\subsection{Lyman alpha emission}

The luminosity of Ly$\alpha$ emission at a frequency $\nu$ is
\be
L_{\nu}^{\rm Ly\alpha} = h\nu_{\rm Ly\alpha}(\epsilon_{\rm Ly\alpha}^{\rm rec}+\epsilon_{\rm Ly\alpha}^{\rm coll})\phi(\nu_{\rm Ly\alpha}-\nu,z)V(M_*),
\ee
where $\nu_{\rm Ly\alpha}$ is the frequency of Ly$\alpha$ photons, 
and $V(M_*)$ is the emission volume that can be estimated by Eq.~(\ref{eq:Vneb}) and Eq.~(\ref{eq:VIGM}) 
for the stellar nebulae and the IGM respectively. 
Here $\epsilon_{\rm Ly\alpha}^{\rm rec}$ is the Ly$\alpha$ recombination emission 
rate per $\rm cm^{3}$, which is given by 
$$\epsilon_{\rm Ly\alpha}^{\rm rec}=f_{\rm Ly\alpha}^{\rm rec}n_en_{\rm HII}\alpha_{\rm B}^{\rm rec},$$
where $n_e$ is the electron
number density, $n_{\rm HII}$ is the HII number density, $\alpha_{\rm B}^{\rm rec}$
is the hydrogen case B recombination coefficient, and $f_{\rm Ly\alpha}^{\rm rec}$
is the fraction of the Ly$\alpha$ photons produced in the case B recombination.
This fraction can be estimated through the fitting formula \citep{Cantalupo08}
\be
f_{\rm Ly\alpha}^{\rm rec}(T) = 0.686-0.106{\rm log_{10}}(T_4)-0.009T_4^{-0.44},
\ee
where $T_4=T/10^4$K, which is accurate to 0.1\% for $100<T<10^5{\rm K}$. 
Note that this fraction is actually not sensitive to the temperature, so it can 
be treated as a constant $\sim 0.68$ in the most cases.
The hydrogen case B recombination coefficient $\alpha_{\rm B}^{\rm rec}$ we use here
is from \cite{Hummer94}, which is fitted by \cite{Seager99} as
\be
\alpha_{\rm B}^{\rm rec}(T)=10^{-13}\frac{aT_4^b}{1+cT_4^d}\ \rm [cm^3/s],
\ee
where $a=4.309$, $b=-0.6166$, $c=0.6703$ and $d=0.5300$. We assume a gas temperature of
$T_{\rm gas}=3\times 10^4$ K in our calculation to obtain $\alpha_{\rm B}^{\rm rec}$.

The $\epsilon_{\rm Ly\alpha}^{\rm coll}$ is the collisional emission rate per $\rm cm^3$ given by
\be
\epsilon_{\rm Ly\alpha}^{\rm coll} = C_{\rm Ly\alpha}^{\rm eff}n_en_{\rm HI},
\ee
where $n_{\rm HI}$ is the neutral hydrogen number density and $C_{\rm Ly\alpha}^{\rm eff}$
is the effective collisional excitation coefficient. It has the form \citep{Cantalupo08}
\be
C_{\rm Ly\alpha}^{\rm eff} = C_{1,2p}+C_{1,3s}+C_{1,3d} \, .
\ee
Here we take into account the excitation up to energy level $n=3$ to
produce  Ly-$\alpha$ photons. The higher level emission can be neglected given
the high temperature we consider for this calculation. 
The excitation collisional rate $C_{\rm l,u}$, in cm$^3$ per second, 
can be  written as
\be
C_{l,u} = \frac{8.629\times 10^{-6}}{g_{l}\sqrt{T}}\gamma_{l,u}e^{-E_{\rm l,u}/kT}\ \rm [cm^3/s],
\ee
where $E_{\rm l,u}$ is the energy difference between lower level $l$ and higher 
level $u$, $g_{l}$ is the statistic weight for level $l$, and $\gamma_{l,u}(T)$
is the effective collision strength calculated using the fitting formulae 
from \cite{Giovanardi87}.

The $\phi(\nu_{\rm Ly\alpha}-\nu,z)$ is the Ly$\alpha$ line profile, and we use the
result from \cite{Santos02} where they fitted the simulated Ly$\alpha$ line profile
of \cite{Loeb99} for a homogeneous and expanding IGM:
\ba
\phi(\nu_{\rm Ly\alpha}-\nu,z) = \left\{\begin{array}{ll}
\nu_*(z)d\nu^{-2}{\rm exp}[-\nu_*(z)/d\nu]\ & \textrm{if}\ d\nu>0\\ \nonumber
0\ & \textrm{if}\ d\nu\le0 \end{array}\ ,\right.
\ea
where $d\nu=\nu_{\rm Ly\alpha}-\nu$, and 
$$\nu_*(z)=1.5\times10^{11}{\rm Hz}\ \left(\frac{\Omega_{\rm b}h^2}{0.019}\right)\left(\frac{h}{0.7}\right)^{-1}\frac{(1+z)^3}{E(z)}.$$
Here $E(z)=H(z)/H_0$, and we assume the flat $\Lambda$CDM model and take 
$E(z)=\sqrt{\Omega_{\rm m}(1+z)^3+\Omega_{\Lambda}}$.

\subsection{Free-free and free-bound emission}

For free-free and free-bound emission we again follow the same approach as Fernandez \& Komatsu (2006).
Following their derivation, the continuum luminosity of these two processes at frequency $\nu$ has
the same form with
\be
L_{\nu}^{\rm ff,fb} = 4\pi j_{\nu}^{\rm ff,fb} V(M_*),
\ee
where $j_{\nu}^{\rm ff,fb}$ is the specific emission coefficient for free-free and
free-bound emission \citep{Dopita02}
\be
j_{\nu}^{\rm ff,fb} = 5.44\times 10^{-39}\frac{e^{-h\nu/kT}}{\sqrt{T}}n_en_pg_{\rm eff}^{\rm ff,fb}\ \ \rm [erg/cm^3/s/Hz/sr].
\ee
Here $n_p$ is the proton number density, $T$ is the gas temperature, and 
$g_{\rm eff}^{\rm ff,fb}$ is effective Gaunt factor for free-free and free-bound
emission, which takes the form as
\ba
g_{\rm eff}^{\rm ff,fb} = \left\{\begin{array}{ll}
\bar{g}_{\rm ff}\ & \textrm{free-free}\\
\frac{x_ne^{x_n}}{n}g_{\rm fb}(n)\ & \textrm{free-bound}\end{array}\ ,\right.
\ea
where $\bar{g}_{\rm ff}\simeq 1.1$ is the thermal averaged Gaunt factor of free-free 
emission and $g_{\rm fb}(n)\simeq 1.05$ is the free-bound emission Gaunt factor for a
different energy level $n$. These values have an accuracy of 10\% \citep{Karzas61}. 
In above $x_n=R_y/(kT_gn^2)$, where $Ry/kT_g$ is around 10 for the parameter space we are interested 
in \citep{Fernandez06}. The energy level $n$ is determined by the emission photon frequency $\nu$.
If $cR_y/n'^2<\nu<cR_y/(n'-1)^2$, and then $n=n'$ where $R_y=1.1\times 10^7\ \rm m^{-1}$ is the 
Rydberg constant. Note that the $n$ here starts at $n=2$, since the photons from 
$n=1$ can be easily absorbed by other neutral hydrogen atoms and be ionized instantly.

\subsection{Two-photon emission}

For the two-photon process we also follow the approach of Fernandez \& Komatsu (2006) and write the
luminosity as
\be
L_{\nu}^{\rm 2ph} = \frac{2h\nu}{\nu_{\rm Ly\alpha}}P(\nu/\nu_{\rm Ly\alpha})\epsilon_{\rm 2ph}V(M_*),
\ee
where $\epsilon_{\rm 2ph}=f_{\rm 2ph}n_en_{\rm HII}\alpha_{\rm B}^{\rm rec}$ is
the two-photon emission rate per $\rm cm^3$, and 
$f_{\rm 2ph}\simeq(1-f_{\rm Ly\alpha}^{\rm rec})$.
The $P(\nu/\nu_{\rm Ly\alpha})d\nu/\nu_{\rm Ly\alpha}$ is the normalized probability
of generating one photon in the range $d\nu/\nu_{\rm Ly\alpha}$ 
from per two-photon decay. We use the fitting formula derived in \cite{Fernandez06}
\ba
P(y) &=& 1.307-2.627(y-0.5)^2 \nonumber \\
     &+& 2.563(y-0.5)^4-51.69(y-0.5)^6 \nonumber
\ea

where $y=d\nu/\nu_{\rm Ly\alpha}$, 
which is a good fit to the data given in \cite{Brown70}.

\subsection{Luminosity mass density and total emission}

Then following Fernandez \& Komatsu (2006) and Fernandez et al. (2010), we can derive the mean luminosity mass density for each 
emission source by integrating over the IMF for the Pop II or Pop III stars
\be
l_{\nu} = \frac{\int dM_* f(M_*) L_{\nu}(M_*)}{\langle M_* \rangle},
\ee
where the ranges of the integral are from 3 to 150 $M_{\sun}$ with the IMF given by Salpeter (1955) for Pop II stars,
and from 3 to 500 $M_{\sun}$ with the IMF in Larson (1998) for Pop III stars. 
Here, the average mass  
$\langle M_*\rangle$ 
is given as
\be 
\langle M_*\rangle = \int_0^\infty dM_*\ M_*f(M_*) \, ,
\ee
where $f(M_*)$ is the normalized  IMF with $$\int_0^\infty dM_*f(M_*)=1.$$ 

Note that this expression is only
valid when the main sequence lifetime is larger than the star formation time scale. Otherwise, it should be evaluated by
\be \label{eq:lv_t_SF}
l_{\nu} = \frac{\int dM_* f(M_*) L_{\nu}(M_*)\tau_*(M_*)}{t_{\rm SF}(z)\langle M_* \rangle}.
\ee 
Here $t_{\rm SF}$ is the star formation time scale which is given by
\be \label{eq:t_SF}
t_{\rm SF}(z) = \left(\frac{d{\rm ln}\rho_*(z)}{dt}\right)^{-1},
\ee
where $\rho(z)$ is the stellar mass density at $z$, which is related with the comoving star formation rate
density (SFRD) as $\psi(z)=d\rho_*(z)/dt$. We use the halo mass function to calculate $\psi(z)$ and
$\rho_*(z)$ and the details are described in the next Section. We note that the $t_{\rm SF}$ is
important for the estimation of the $l_{\nu}$ as discussed in \cite{Fernandez10}. 

Finally we obtain the total luminosity mass density from the stellar nebulae
\be
l_{\nu}^{\rm neb} = l_{\nu}^*+(1-f_{\rm esc})(l_{\nu}^{\rm Ly\alpha}+l_{\nu}^{\rm ff}+l_{\nu}^{\rm fb}+l_{\nu}^{\rm 2ph})
\ee
and the same from the IGM
\be
l_{\nu}^{\rm IGM} = f_{\rm esc}(l_{\nu}^{\rm Ly\alpha}+l_{\nu}^{\rm ff}+l_{\nu}^{\rm fb}+l_{\nu}^{\rm 2ph}),
\ee
where $f_{\rm esc}$ is the escape fraction of the ionization photons that propagate into the IGM from the nebulae 
surrounding the stars.

In Fig.~\ref{fig:Lv}, we show the total luminosity mass density as a function of the rest-frame 
wavelength $\lambda$ from the stellar nebula and the IGM for Pop II and Pop III stars respectively. 
The contributions from the different sources we consider are also shown for the stellar 
nebula case. Here we set $f_*=0.03$ and $f_{\rm esc}=0.5$ for both Pop II and Pop III cases.

With these parameters we find that the stellar
spectrum is dominant for Pop II stars while the ``background" spectrum, such as Ly-$\alpha$ and
free-bound, are comparable with or even larger than the stellar spectrum for Pop III stars.
Also, as can be seen, the $l_{\nu}$ from the IGM is much lower than that from the stellar
nebula for both Pop II and Pop III cases, and the total $l_{\nu}$ from the Pop III stars is
similar to that from the Pop II stars. These results are already discussed in Fernandez \& Komatsu (2006).

\section{Reionization History and UV Luminosity Density}

\begin{figure*}[htb]
\includegraphics[scale = 0.4]{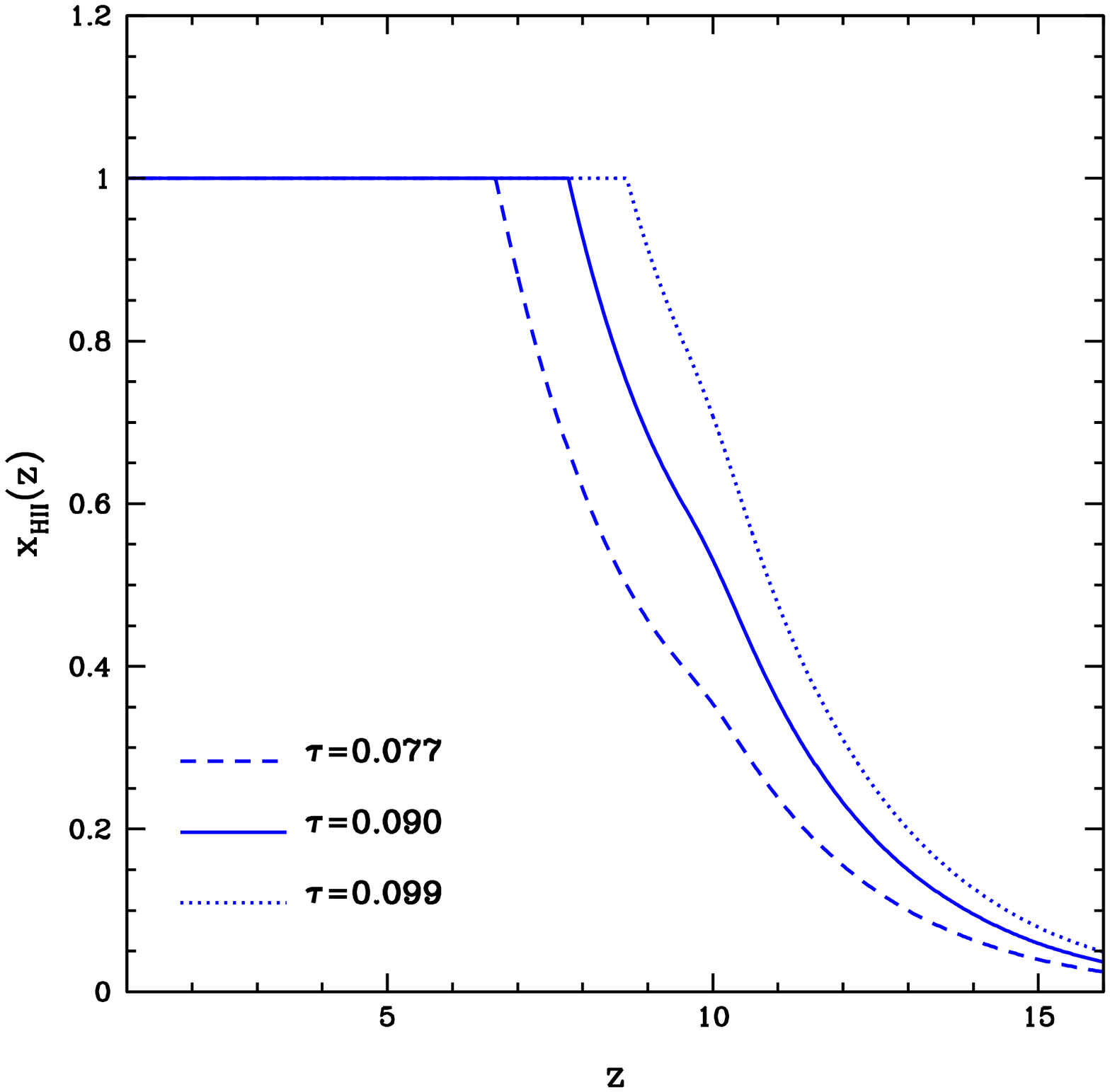}
\includegraphics[scale = 0.4]{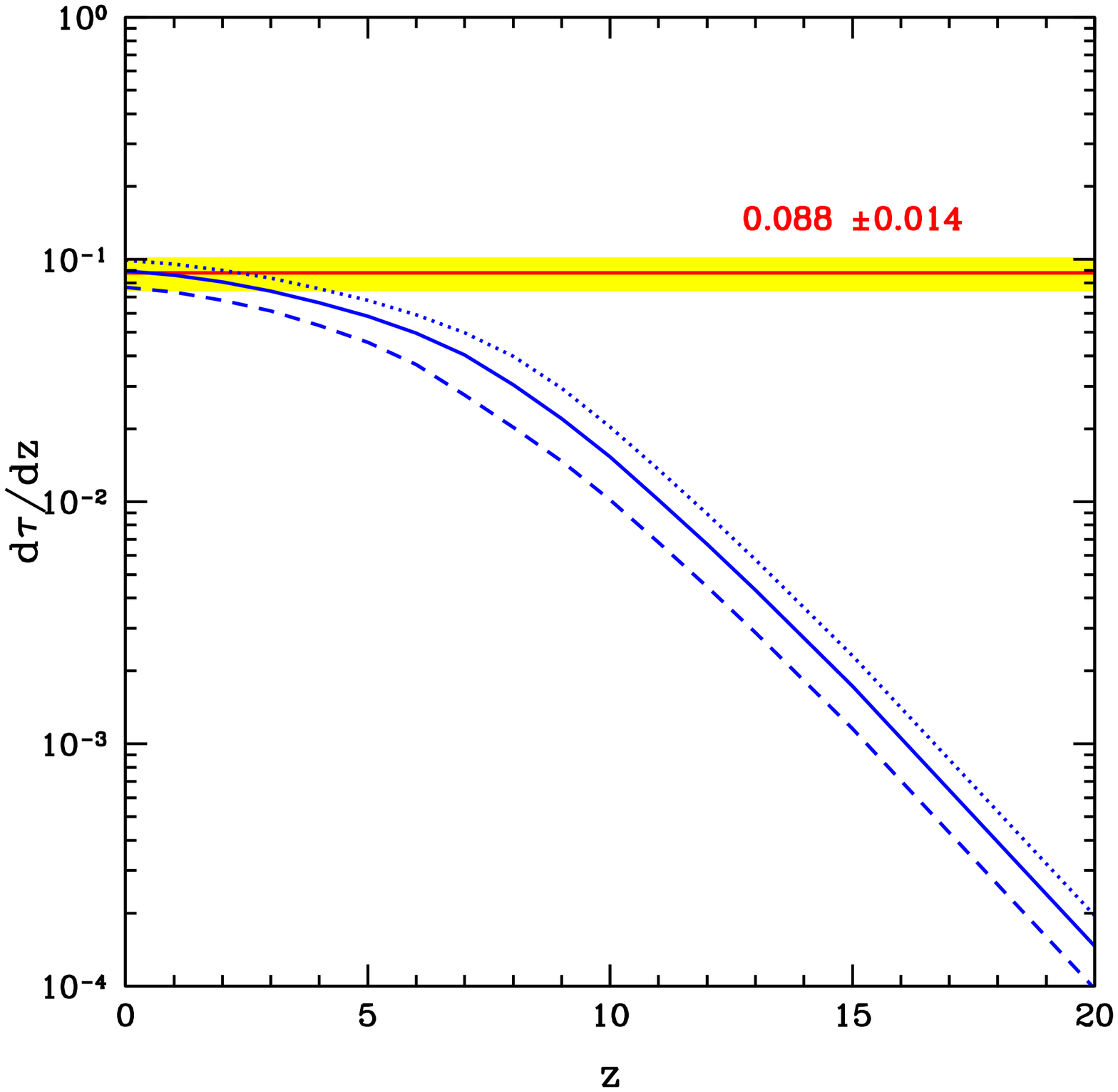}
\caption{\label{fig:xe} 
{\it Left:} The hydrogen reionization fraction $x_{\rm HII}$ as a function of redshift $z$ for
three cases of the $f_*=0.02$, $0.03$ and $0.04$. The solid line with $f_*=0.03$ indicates the
optical depth $\tau = 0.090$ which is close to the
result of WMAP 7-year which gives $\tau=0.088 \pm 0.014$.
The dashed and dotted lines are derived with $f_*=0.02$ and $0.04$ which denote $\tau=0.077$ and $0.099$, respectively. 
{\it Right:} The dependence of the optical depth on the minimum redshift in Eq.~(\ref{eq:tau}).
The blue solid, dashed and dotted lines are for the three stellar models with $f_*=0.03$, 0.02 and 0.04.
The result of WMAP 7-year is also shown in red solid line with yellow $1\sigma$ error region for comparison.
}
\end{figure*}

To test if the reionization history associated with our stellar model is consistent with that of the current
observations, such as WMAP 7-year results \citep{Komatsu11}, we need to calculate the hydrogen reionization fraction 
$x_{\rm HII}(z)$ as a function of redshift. Following Madau et al. (1998), $x_{\rm HII}(z)$ can be  estimated as
\be \label{eq:xe}
\frac{dx_{\rm HII}}{dt} = \frac{f_{\rm esc}\psi(z)q(z)}{\bar{n}_{\rm H}(z)} - \frac{x_{\rm HII}}{\bar{t}_{\rm rec}},
\ee
where $z_{\rm i}$ is the redshift of the beginning of the reionization epoch (we take
$z_{\rm i}=30$), $\psi(z)$ is the comoving star formation rate density (SFRD),
and the function $q(z)$ is defined as $q(z)\equiv\left(\overline{Q}_{\rm HI}/\langle M_*\rangle\right)\langle \tau_*\rangle$.
Here $\langle \tau_*\rangle$ is the average stellar lifetime which is given by $\langle \tau_*\rangle=\int_0^\infty dM_*\ \tau_*(M_*)f(M_*)$,
and $t_{\rm rec}$ is the volume averaged recombination time, which can be written as
\be
\bar{t}_{\rm rec} = \left[C_{\rm HII}(z)\alpha_{\rm B}^{\rm rec}\bar{n}_{\rm H}(z)(1+Y/4X)\right]^{-1},
\ee
where $C_{\rm HII}\equiv \langle n^2_{\rm HII}\rangle/\langle n_{\rm HII}\rangle^2$ is the clumping factor of
ionized hydrogen. Here we adopt the simulation result from Trac \& Cen (2007). $X=0.75$ and $Y=0.25$ are 
the mass fractions of hydrogen and helium, respectively. Note that we have already considered the escape fraction
$f_{\rm esc}$, so the $C_{\rm HII}$ here is the clumping factor excluding the halos with star formation.

For the comoving SFRD we consider the ongoing star formation model \citep{Santos02}:
\be \label{eq:psi}
\psi(z) = f_*\frac{\Omega_{\rm b}}{\Omega_{\rm m}}\frac{\rm d}{{\rm d}t}\int_{M_{\rm min}}^{\infty}dM\ M\frac{dn}{dM}(M,z),
\ee
where $f_*$ is the star formation efficiency which denotes the fraction of baryons converted to
stars, $dn/dM$ is the halo mass function \citep{Sheth99}, and $M_{\rm min}$
is the threshold mass for a dark matter halo to form a galaxy during reionization. This minimum mass is taken to be a free parameter
and is varied to fit a combination of the WMAP 7-year optical depth and the galaxy LF, as we discuss in Section~6.

We also need a stellar population
evolution model to describe $q(z)$ with the relative fraction of the Pop II and Pop III stars at different
redshifts. In principle there should be a cutoff at some redshift for Pop III stars as they are not expected to form 
at low redshifts once the gas is polluted by metals. We assume this cutoff is not lower than $z=6$ when the universe is fully ionized.
We use the error function to denote the population fraction as
\be \label{eq:fP}
f_{\rm P}(z) = \frac{1}{2}\left[ 1 + {\rm erf}\left( \frac{z_{\rm t}-10}{\sigma_{\rm P}} \right) \right],
\ee
where $\sigma_{\rm P}=0.5$ is the population transition width. Then the term 
$q(z)$ in Eq.~(\ref{eq:xe}) can be expressed by
\be
q(z) = f_{\rm P}\frac{\overline{Q}_{\rm HI}^{\rm Pop III}}{\langle M^{\rm Pop III}_*\rangle}\langle \tau_*^{\rm PopIII}\rangle + (1-f_{\rm P})\frac{\overline{Q}_{\rm HI}^{\rm Pop II}}{\langle M^{\rm Pop II}_*\rangle}\langle \tau_*^{\rm Pop II}\rangle.
\ee
As our default model we assume that the Pop III stars are mainly dominant for $z\gtrsim 10$ while Pop II
stars are for $z\lesssim 10$ with $z_{\rm t}=10$. When we present our results we also show results for three additional values of the transition from
$z_{\rm t}=10$ to 30. 

We estimate the optical depth to electron scattering with the reionization fraction
$x_e(z)$ as
\be \label{eq:tau}
\tau = \int_{0}^{\infty} dz \frac{c}{H(z)} \frac{n_{e}(z)\sigma_{\rm T}}{1+z},
\ee
where $\sigma_{\rm T} = 6.65\times 10^{-29}\ \rm m^2$ is the Thompson scattering cross-section,
and $n_{e}(z)=x_{\rm HII}(z)\bar{n}_{\rm H}(z)(1+\eta Y/4X)$ is the electron number density of the Universe
at redshift $z$, and we assume the helium is singly ionized for $z>4$ ($\eta=1$) and doubly ionized
for $z<4$ ($\eta=2$) \citep{Kuhlen12}.

In Fig.~\ref{fig:xe}, we plot the hydrogen reionization fraction $x_{\rm HII}$ vs. $z$ for three optical depth $\tau$
with three stellar models. The blue solid line denotes the $\tau=0.090$ with $f_*=0.03$ which is close to the result
of WMAP 7-year data with $\tau=0.088 \pm 0.014$ \citep{Komatsu11}. 
In this case reionization ends around $z_{\rm e}\simeq8$ which is consistent
with the current studies, and we find Pop III stars can ionize the Universe to $\sim 60\%$ by
$z=10$ and Pop II stars is responsible for the rest of the reionization over an interval $\Delta z\simeq 2$.

For $f_*=0.02$ and 0.04, we find $\tau=0.077$, shown with a dashed line and $\tau=0.099$, shown with a dotted line, respectively.
These two reionization histories are such that $z_{\rm e}\sim 6.5$ and 9, respectively.
For case with $\tau=0.077$, the Pop III stars ionize $\sim 40\%$ of the Universe and the 
Pop II stars are needed over the interval $\Delta z\simeq 3.5$ to complete reionization.
For the third case with $\tau=0.099$, Pop III stars ionize $80\%$ of the Universe at $z=10$ with
Pop II stars completing the rest $20\%$ in an interval of $\Delta z\simeq 1$

In Fig.~\ref{fig:xe}, we also show the dependence of the optical depth on the minimum redshift
$z_{\rm min}$ in Eq.~(\ref{eq:tau}). The blue solid, dashed and dotted lines are for the three stellar 
models with $\tau=0.090$, 0.077 and 0.099 for $z_{\rm min}=0$. The WMAP 7-year result with $\tau=0.088\pm 0.014$ 
is also shown with a red solid line and a yellow $1\sigma$ error region. We find the slope of the curves is steeper for 
$z_{\rm min}>10$ and flatter for $z_{\rm min}<10$ which is caused by the Pop III to Pop II transition around $z=10$ in our model.
Note that this transition is arbitrarily chosen. We varied the transition redshift and also cases where PopII and PopIII stars are mixed in
with different fractions at  different redshifts. In all these cases we found results that are generally consistent with each other. Thus, the three choices related to the reionization history that
we show here to keep this presentation simple are not biased with respect to the final result related to the IR background intensity that we are trying to estimate in this paper.

\begin{figure}[htb]
\includegraphics[scale = 0.4]{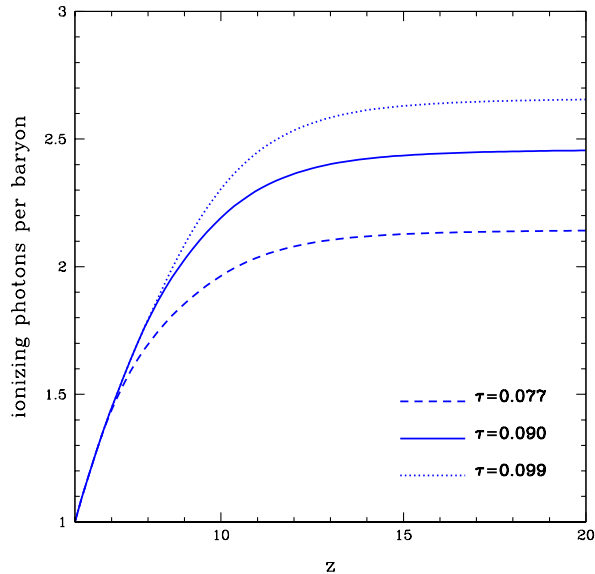}
\caption{\label{fig:ph_bar} The total number of ionizing photons per baryon to 
maintain the ionized IGM between the redshift of the end of the reionization and $z$.
The blue dotted, solid and dashed lines are derived from the model with $f_*=0.02$, $0.03$
and $0.04$, respectively.
}
\end{figure}

We also estimate the total number of ionizing photons per baryon required to maintain the ionized IGM between
$z_{\rm end}$ and z,
\be
N_{\rm ion}^{p}(z) = N_{\rm end} + \int_z^{z_{\rm end}} \frac{x_{\rm HII}(z')}{t_{\rm rec}}\frac{dt}{dz'}dz',
\ee
where $z_{\rm end}=6$ is the redshift of the end of the reionization, and if we assume the helium is also totally 
singly ionized at $z_{\rm end}=6$ we can get $N_{\rm end}\simeq1$. In fig.~\ref{fig:ph_bar}, we show the 
$N_{\rm ion}^{p}(z)$ at different $z$ for three $f_*$ cases. We find the number keeps going up until around
$z=15$ and becomes constant $\sim 2.5$ at higher redshift for three cases of the reionization histories.

\section{The near-IR EBL Intensity from Reionization}

\begin{figure*}[htb]
\centerline{
\includegraphics[scale = 0.4]{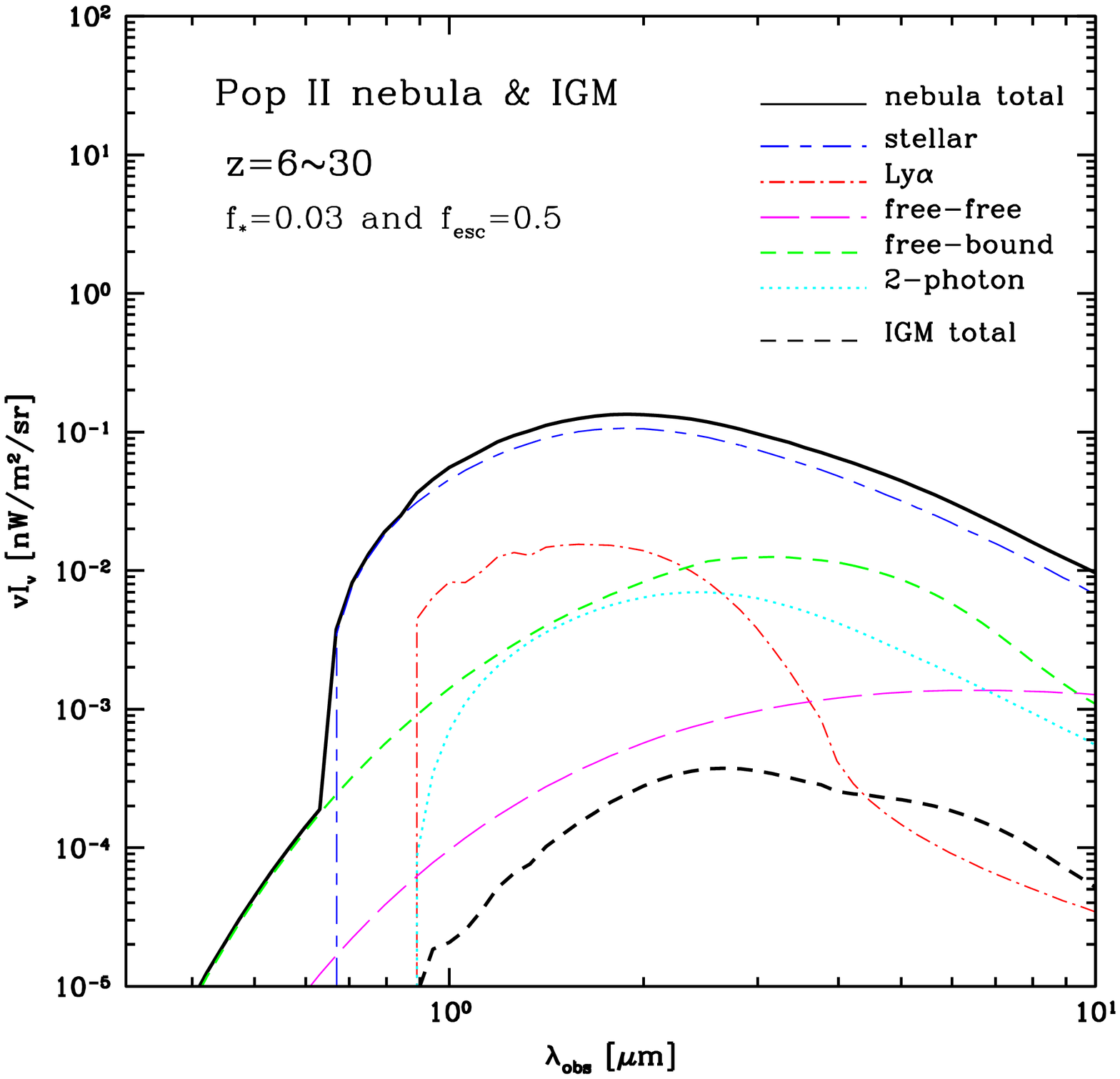}
\includegraphics[scale = 0.4]{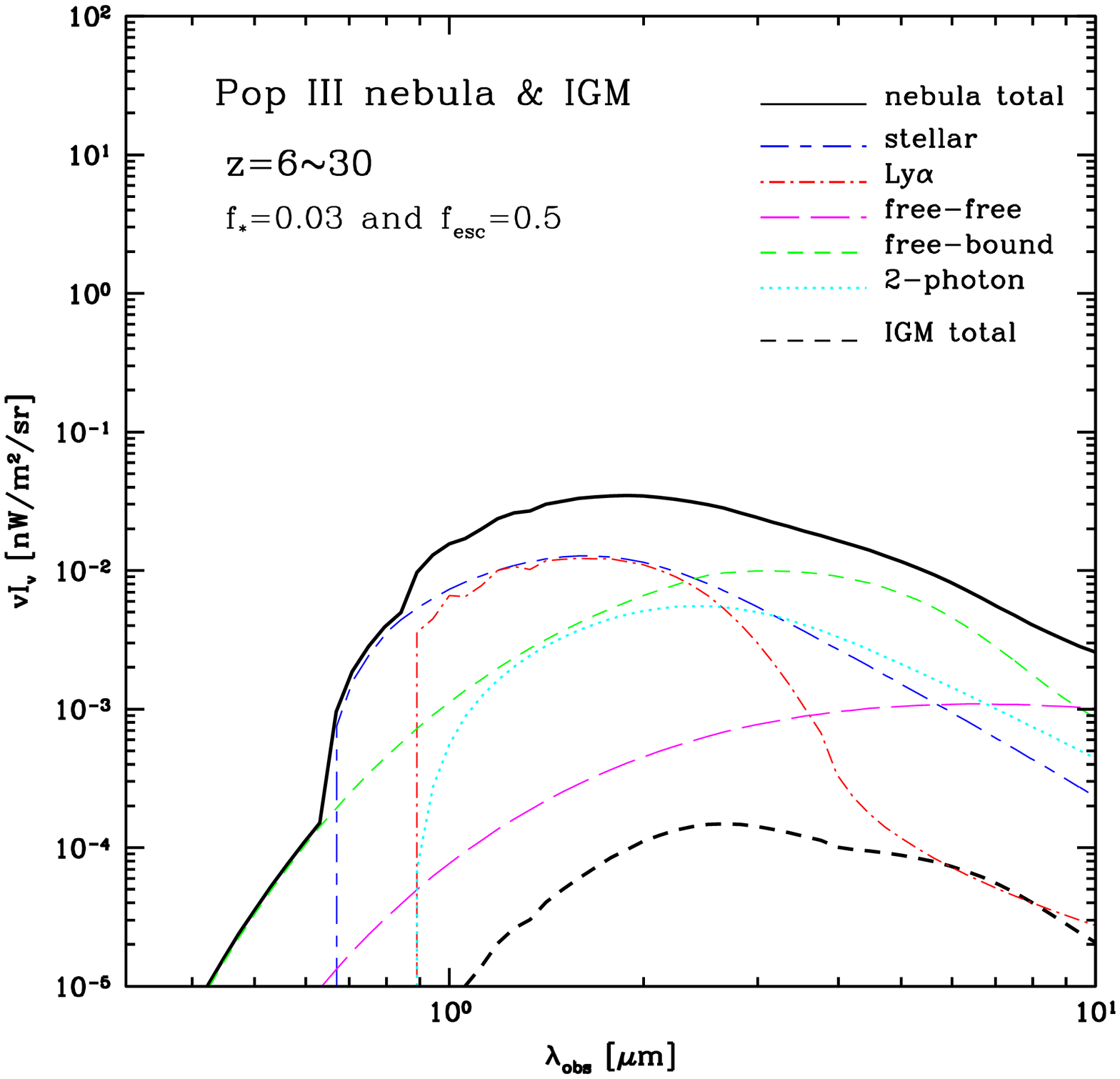}
}
\caption{\label{fig:Iv} The spectrum of the near-IR EBL intensity vs. the observed frame wavelength 
$\lambda_{\rm obs}$ for the Pop II and Pop III stars integrated from $z=6$ to $30$. 
We take $f_*=0.03$ and $f_{\rm esc}=0.5$ for both Pop II and Pop III cases. We find the spectrum 
of the stellar nebula is much larger than that from the IGM for both Pop II and Pop III stars 
which is similar with what is shown in Fig.~\ref{fig:Lv}. However, unlike the luminosity 
density case, the spectrum from the Pop II stars is larger than that from the Pop III stars.
This is basically because the typical lifetime of the Pop II stars is larger than that of 
Pop III stars.}
\end{figure*}

The mean cosmic infrared background can be estimated by
\be \label{eq:Iv}
\nu_{\rm obs}\bar{I}_{\nu_{\rm obs}}=\int_{z_{\rm min}}^{z_{\rm max}}dz\frac{c}{H(z)}\frac{\nu(z)\bar{j}_{\nu}(z)}{(1+z)^2},
\ee
where $\nu(z)=(1+z)\nu_{\rm obs}$, and we take $z_{\rm min}=6$ and $z_{\rm max}=30$
in the calculation. This redshift range can fully take account of the emission from
the Pop III and early Pop II stars, which is redshifted into the near-IR band.
We take this to be the form of
\be \label{eq:jv_p}
\bar{j}_{\nu}(z) = f_{\rm P}\bar{j}_{\nu}^{\rm Pop III}(z) + (1-f_{\rm P})\bar{j}_{\nu}^{\rm Pop II}(z),
\ee
where $\bar{j}_{\nu}^{\rm Pop III}$ and $\bar{j}_{\nu}^{\rm Pop II}$ are the 
 comoving specific emission coefficients
\be\label{eq:jv}
\bar{j}_{\nu}^i(z) = \frac{1}{4\pi}l^i_{\nu}\langle \tau^i_* \rangle\psi(z),
\ee
where $l^i_{\nu}$ is the luminosity mass density at $\nu$, $\langle \tau^i_* \rangle$ is the mean stellar lifetime of each of the stellar type
and $\psi(z)$ is the comoving SFRD given by Eq.~(\ref{eq:psi}). 

In Fig.~\ref{fig:Iv}, we show the spectrum $I_{\nu}$ of the near-IR background light intensity from both Pop II and Pop III stars. 
We assume that the reionization is ending around $z=6$ and integrate up to $z=30$ to determine $I_{\nu}$.
Similar to Fig.~\ref{fig:Lv}, we plot the total spectrum from the stellar nebula and the IGM
for both Pop II and Pop III stars. The contributions from different sources we consider
are also shown in colored thin lines. 

Here we still take the same value for $f_*$ and $f_{\rm esc}$ as in Fig.~1 related to the $l_{\nu}$ calculation.
Similar to the luminosity mass density, the spectrum from stellar nebula is much larger than that from the IGM, and the stellar 
spectrum is higher for Pop II stars while  the ``background" spectrum is higher for Pop III stars.
However, different from the luminosity mass density, we now find the spectrum from Pop II stars is larger 
than that from Pop III stars. This is basically because the typical lifetime of Pop II stars is 
longer than that of the Pop III stars.

\section{Angular power spectrum}

The angular cross power spectrum of the infrared emission at observed frequencies $\nu$ and $\nu'$
for a multipole $\ell$ is
\be \label{eq:Cl}
C_{\ell}^{\nu\nu'} = \int_{z_{\rm min}}^{z_{\rm max}} dz\left(\frac{d\chi}{dz}\right)\left(\frac{a}{\chi}\right)^2\bar{j}_{\nu}(z)\bar{j}_{\nu'}(z)P_{\rm gg}(k,z),
\ee
where $\chi$ is the comoving angular diameter distance, $a=(1+z)^{-1}$ is the scale factor,
and $\bar{j}_{\nu}(z)$ is the mean emission per comoving volume at frequency $\nu$ and 
redshift $z$. If we just take account of the flux lower than a upper cut $S_{\rm cut}$, 
and then $\bar{j}_{\nu}(z)$ can be written as
\be
\bar{j}_{\nu}(z) = (1+z)\int_0^{S_{\rm cut}} dS\ S\frac{d^2N}{dSdz}.
\ee
Here $S$ is the source flux and $N$ is the number of sources. This quantity is 
just the comoving specific emission coefficient we derive in the last section.
The $P_{\rm gg}(k,z)$ is the galaxy power spectrum at wavenumber $k=\ell/\chi$ 
and redshift $z$, and we will make use of the model of halo occupation distribution
to calculate the $P_{\rm gg}(k,z)$.

\subsection{First Galaxy Clustering}

To calculate the $P_{\rm gg}$ we extend the linear theory model of Cooray et al. (2004) and make use of the 
halo occupation distribution (HOD) for first-light galaxies during reionization. The galaxy power spectrum can be written as
\be
P_{\rm gg}(k,z) = P^{\rm 1h}_{\rm gg}(k,z)+P^{\rm 2h}_{\rm gg}(k,z),
\ee
Where $P^{\rm 1h}_{\rm gg}$ and $P^{\rm 2h}_{\rm gg}$ denote the power spectrum
contributed by galaxies in a single dark matter halo and galaxies in two different
dark matter halos respectively. Then we can write \citep{Cooray02}
\ba
P^{\rm 1h}_{\rm gg} &=& \int dM\ \frac{dn}{dM}\frac{\langle N_{\rm gal}(N_{\rm gal}-1)\rangle}{\bar{n}^2_{\rm gal}} u^p(M,k),\\
P^{\rm 2h}_{\rm gg} &=& P_{\rm lin}\left[\int dM\ b(M,z)\frac{dn}{dM}\frac{\langle N_{\rm gal}\rangle}{\bar{n}_{\rm gal}}u(M,k)\right]^2.
\ea
Here $M$ is the halo mass, $dn/dM(M,z)$ is the halo mass function, $u(M,k)$ is the
Fourier transform of the NFW halo density profile \citep{Navarro95}, $p=1$ when 
$\langle N_{\rm gal}(N_{\rm gal}-1)\rangle \le 1$ and $p=2$ otherwise \citep{Cooray02}, 
$b(M,z)$ is the halo bias \citep{Sheth99}, and $P_{\rm lin}(k,z)$ is the linear matter power spectrum \citep{Eisenstein97}. 
The $\bar{n}_{\rm gal}$ is the mean number density of galaxies, which is given by
\be \label{eq:ngal}
\bar{n}_{\rm gal}(z) = \int\ dM \frac{dn}{dM}\langle N_{\rm gal}\rangle.
\ee
The $\langle N_{\rm gal}\rangle$ is the mean number of galaxies in a halo with mass $M$,
which is the sum of number of central galaxies and satellite galaxies \citep{Zheng05}
\be \label{eq:Ngal}
\langle N_{\rm gal}\rangle = \langle N_{\rm cen}\rangle + \langle N_{\rm sat}\rangle,
\ee
where we define
\be
\langle N_{\rm cen}\rangle = \frac{1}{2}\left[ 1+{\rm erf}\left( \frac{{\rm log}M-{\rm log}M_{\rm min}}{\sigma_{{\rm log}M}} \right) \right],
\ee
and
\be
\langle N_{\rm sat}\rangle = \frac{1}{2}\left[ 1+{\rm erf}\left( \frac{{\rm log}M-{\rm log}M_0}{\sigma_{{\rm log}M}} \right) \right]\left( \frac{M}{M_{\rm sat}} \right)^{\alpha_{\rm s}}.
\ee
In this definition, the $M_{\rm min}$ denotes the mass of a halo that has 50\% probability
to host a central galaxy, and $\sigma_{{\rm log}M}$ is the transition width.
For the satellite galaxies, $M_0$ is the truncation mass for satellites, $M_{\rm sat}$ is the
normalization mass and $\alpha_{\rm s}$ denotes the slope of the power-law relation about the halo mass $M$.
We assume $M_0$ is always larger than $M_{\rm min}$ since there should not be satellites without
central galaxy, and assume $M_0=2M_{\rm min}$. We take $M_{\rm sat}=15M_{\rm min}$, $\sigma_{{\rm log}M}=0.3$,
and $\alpha_{\rm s}=1.5$ in this paper. If assuming a Poisson distribution 
for satellite galaxies, we can get
\be
\langle N_{\rm gal}(N_{\rm gal}-1)\rangle \simeq 2\langle N_{\rm sat}\rangle\langle N_{\rm cen}\rangle+\langle N_{\rm sat}\rangle^2.
\ee
This expression could take account of the case $0<\langle N_{\rm cen}\rangle<1$
and is consistent with our definitions for the $\langle N_{\rm cen}\rangle$ and 
$\langle N_{\rm sat}\rangle$.

\subsection{Poisson Fluctuations}

The clustering measurements are affected by the Poisson fluctuations associated with the
shot-noise caused by the discrete and finite number of galaxies from which clustering is measured.
Assuming a Poisson distribution the $\ell$-independent shot-noise power spectrum is
\be
C_{\ell}^{\rm shot} = \int_0^{S_{\rm cut}} dS\ S^2\frac{dN}{dS}.
\ee

To estimate $C_{\ell}^{\rm shot}$, we first define the luminosity mass density for 
the mass of the dark matter halos at frequency $\nu$ 
$$l_{\nu}^h=\frac{L_{\nu}}{M}\simeq f_*\frac{\Omega_{\rm b}}{\Omega_{\rm m}}l_{\nu}^s,$$
where $l_{\nu}^s=L_{\nu}/M_*$ is the luminosity mass density for the stellar mass discussed
in Section 3. We derive the 3-D shot-noise power 
spectrum by assuming $L_{\nu}$ is proportional to the halo mass $M$, i.e. $l_{\nu}^h$ 
is independent on $M$
\be\label{eq:Pshot}
P_{\nu}^{\rm shot}(z) = \left(\frac{l_{\nu}^h}{4\pi} \right)^2\int dM\ M^2\frac{dn}{dM}(M,z).
\ee
Then the 2-D shot-noise power spectrum can be written as
\be
C_{\ell}^{\nu\nu',\rm shot} = \int_{z_{\rm min}}^{z_{\rm max}}dz\left(\frac{d\chi}{dz}\right)\left(\frac{a}{\chi}\right)^2P_{\nu}^{\rm shot}(z).
\ee

\subsection{Band-Averaged Intensity Power Spectrum}

For a specific near-IR observation with a band frequency from $\nu_1$ to $\nu_2$,
we can define a band-averaged luminosity mass density as
\be
l = \frac{1}{\Delta \nu}\int_{\nu_1(1+z)}^{\nu_2(1+z)} d\nu\ l_{\nu},
\ee
where $\Delta \nu=\nu_2-\nu_1$ is the bandwidth.
Then we can derive the band-averaged comoving specific emission coefficient $\bar{j}(z)$ using
Eq.~(\ref{eq:jv}) and the 3-D shot-noise power spectrum $P^{\rm shot}$ using Eq.~(\ref{eq:Pshot}) 
respectively. 

Finally we find the band-averaged angular cross power spectrum and shot-noise power spectrum are
\be \label{eq:Cl_band}
C_{\ell} = \int_{z_{\rm min}}^{z_{\rm max}} dz\left(\frac{d\chi}{dz}\right)\left(\frac{a^2}{\chi}\right)^2\bar{j}(z)\bar{j}(z)P_{\rm gg}(k,z),
\ee
and
\be
C_{\ell}^{\rm shot} = \int_{z_{\rm min}}^{z_{\rm max}}dz\left(\frac{d\chi}{dz}\right)\left(\frac{a^2}{\chi}\right)^2P^{\rm shot}(z) \, ,
\ee
respectively. Note that here we have a factor $a^4$ instead of $a^2$ in $C_{\ell}^{\nu\nu'}$ and 
$C_{\ell}^{\nu\nu',\rm shot}$ and this dependence has been explained in the Appendix of \citet{Fernandez10}.

\section{Results and Discussion}

In this Section, we first estimate the infrared background intensity and then discuss the angular power
spectrum as derived previously. We also compare our estimation with the observational data and discuss the dependence
of the result on the parameters in the model.

\subsection{Near-IR EBL from Reionization}

\begin{figure*}[htb]
\includegraphics[scale = 0.4]{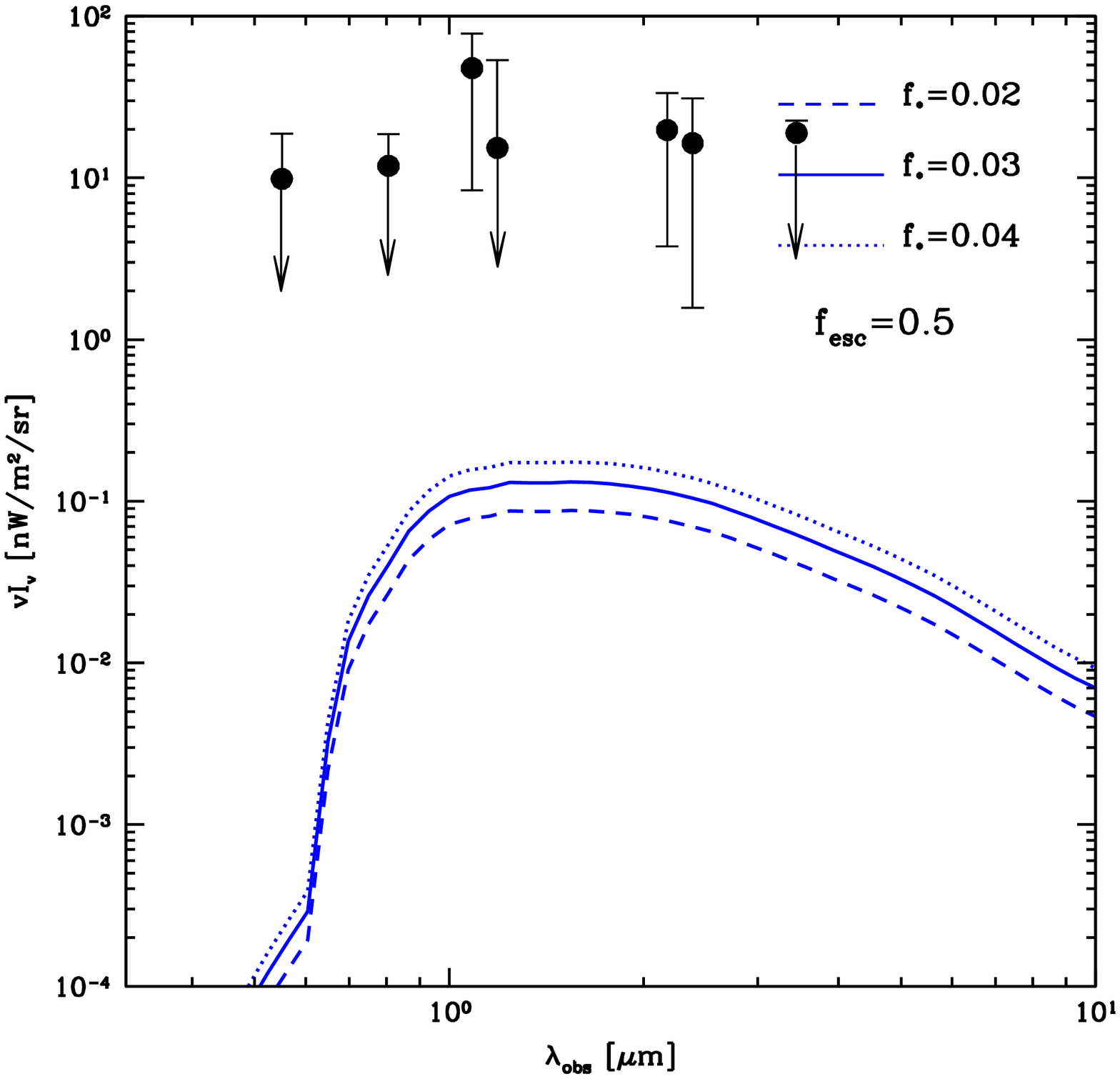}
\includegraphics[scale = 0.4]{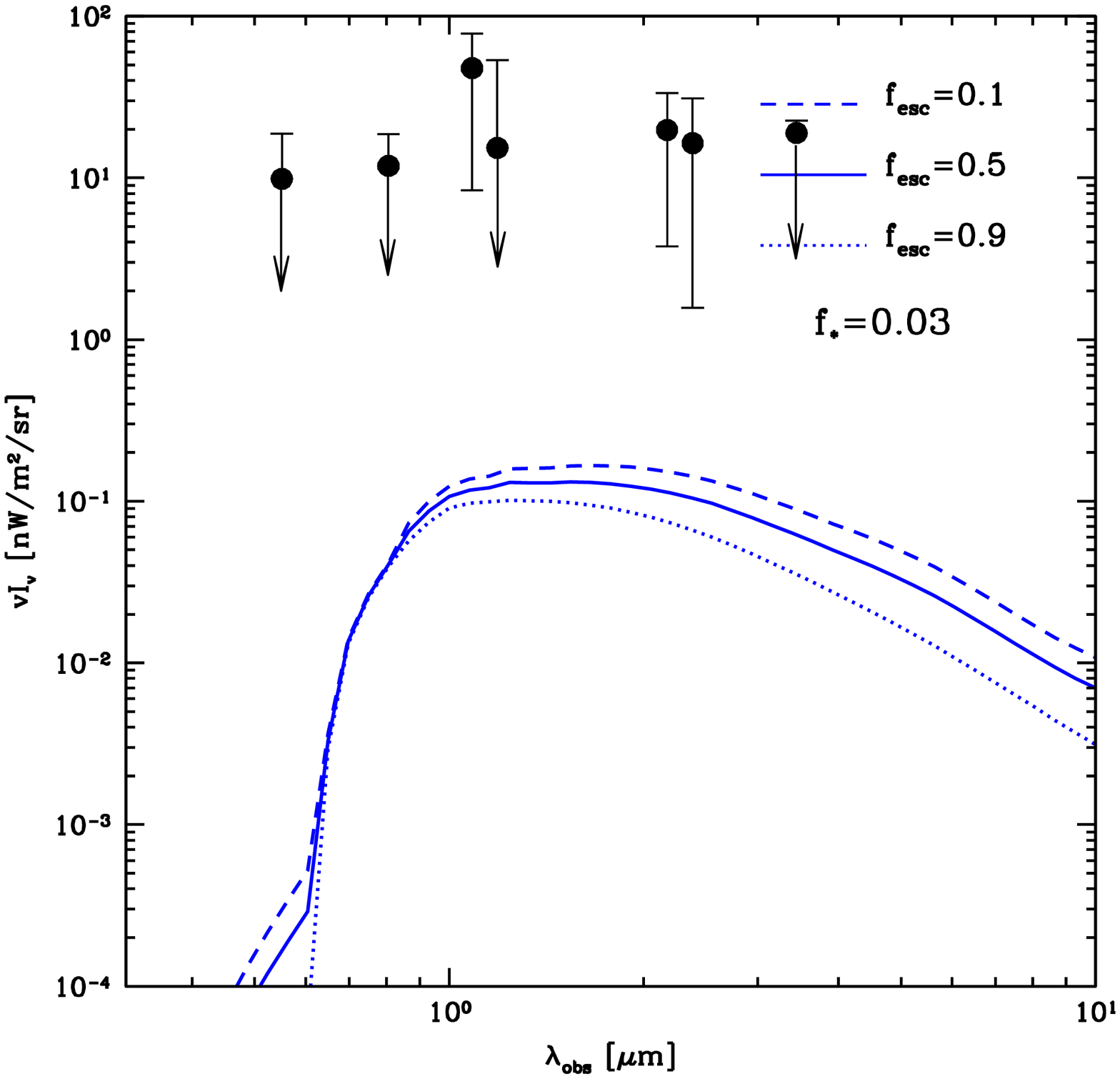}
\caption{\label{fig:Iv_tau} The total near-IR background intensity spectrum for three assumed values of $f_*$ at fixed
$f_{\rm esc}$ ({\it left panel}) and the same with $f_{\rm esc}$ varied at fixed $f_*$ ({\it right panel}).
In the left panel, the dashed, solid and dotted lines denote the three cases of reionization history 
with $\tau=0.077$, 0.090 and 0.099. The models in the right panel also range in $\tau$ between 0.07 and 0.1.
The data points with error bars are from Santos et al. (2002) and show the ``DIRBE excess'' defined as the
difference between total DIRBE background and the integrated galaxy light at each of the wavelengths. The points shown with arrows are strictly upper limits
since the low-end of the error is consistent with zero for them with significance less than 2$\sigma$.
}
\end{figure*}

In Fig.~\ref{fig:Iv_tau}, we show the spectrum of the  near-IR intensity vs. the observed frequency for three
cases of the reionization history with $\tau=0.077$, 0.090 and 0.099 corresponding to $f_*=0.02$, 0.03 and 0.04. 
Here the spectrum is the total spectrum of the sum of that from both stellar nebula and the IGM, which are calculated 
by putting Eq.~(\ref{eq:jv_p}) into Eq.~(\ref{eq:Iv}). Also, we set $f_{\rm esc}=0.5$ for both
Pop II and Pop III stars. The right panel of Fig.~\ref{fig:Iv_tau} shows three cases with $_{\rm esc}$ varied at fixed $f_*$.
In both panels, for comparison, we plot the observational data in terms of the excess of EBL relative to 
the integrated galaxy light (IGL) of known galaxy populations at low redshifts (Madau \& Pozzetti 2000). 
The excess EBL data plotted here are the same as those shown in Santos et al. (2002) and involves measurements 
mainly from DIRBE with various models for zodiacal light and Galactic stellar contribution.

We find the emission from the Pop II stars dominates the spectrum for all of the three cases. As we have just discussed
in the last Section, this is because the lifetime of the Pop II is longer than the Pop III stars. 
In the spectrum, the shorter and longer wavelength parts are mainly contributed by the ``background'' 
spectrum from Pop III stars while the medium part by Pop II stellar spectrum, so if just the Pop III stars 
get longer lifetime, only the ``background'' spectrum can be effectively reinforced in the total spectrum.

In any case, regardless of assumptions related to the stellar type, we find that the EBL intensity from 
reionization is no more than 0.4 nW m$^{-2}$ sr$^{-1}$. Such an intensity is significantly smaller
than the previous predictions that attempted to explain almost all or a significant fraction of the excess EBL 
seen in DIRBE data relative to IGL estimates. An intensity larger than about 2.5 nW m$^{-2}$ sr$^{-1}$ in the J-band could
be in conflict with metal production considerations and the X-ray background (Madau \& Silk 2005), though 
they do not necessarily require high efficiency factors to generate the required star formation (Fernandez \& Komatsu 2006). 
The difference between our calculation and the previous ones is that we primarily require the reionization model to 
generate a reionization history consistent with the WMAP optical depth. This limits the number of H-ionizing photons per baryon during reionization to be less than 3. 
Previous estimates ignored such a constraint and either focused on explaining all of the DIRBE excess (Santos et al. 2002; Salvaterra \& Ferrara 2003) or amplitude of the
measured near-IR background anisotropies (e.g., Fernandez et al. 2010).

\begin{figure}[htb]
\includegraphics[scale = 0.4]{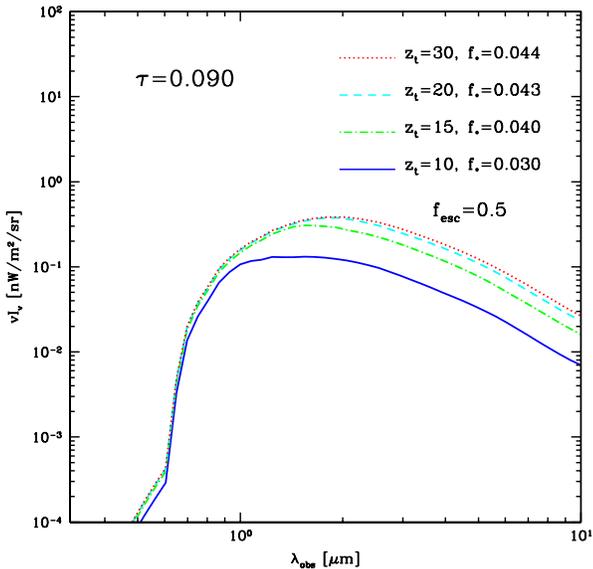}
\caption{\label{fig:Iv_zt} The total near-IR background intensity spectrum 
for transition redshifts $z_{\rm t}$ of Pop II and Pop III stars between 10 and 30. To maintain
the same optical depth to electron scattering of $\tau=0.090$, a higher intensity is predicted as $z_{\rm t}$ is increased.
For example the integrated intensity when $z_{\rm t}=30$ (red dotted line) is $\sim$ 3 to 4 
times greater than when $z_{\rm t}=10$ (blue solid line) for $\lambda_{\rm obs}>2 \rm \mu m$.
}
\end{figure}

In Fig.~\ref{fig:Iv_zt}, we show the total near-IR background intensity spectrum for different 
transition redshifts $z_{\rm t}$ from Pop II to Pop III stars with increasing redshift. Here we take four transition redshifts
$z_{\rm t}=10$, 15, 20 and 30 using Eq.~(\ref{eq:fP}). To maintain the same optical depth to electron scattering $\tau=0.090$,
we find a higher intensity is predicted for higher $z_{\rm t}$. Correspondingly $f_*$ should be increased to values of
0.040, 0.043 and 0.044 for $z_{\rm t}=15$, 20 and 30, compared to $f_*=0.03$ for $z_{\rm t}=10$. 
If $z_{\rm t}=30$ (red dotted line) the integrated intensity is about $\sim$ 3 to 4 times greater 
than the case for $z_{\rm t}=10$ (blue solid line) when $\lambda_{\rm obs}>2 \rm \mu m$.
Also, the peak of the integrated intensity spectrum moves to longer wavelengths as $z_{\rm t}$ is increased.
This is because the Pop III stars are generally hotter than the Pop II stars, which can produce more ionizing
photons. When the transition redshift is higher, the longer the Pop II stars dominate the Universe, 
and less ionizing photons are produced when compared to the case for a low $z_{\rm t}$. To keep the reionization 
history unchanged, we need more Pop II stars to generate enough ionizing photons. This results in a near-IR 
background intensity that is higher.

\subsection{Bright-end Galaxy Luminosity Functions}

\begin{figure*}[htb]
\includegraphics[scale = 0.4]{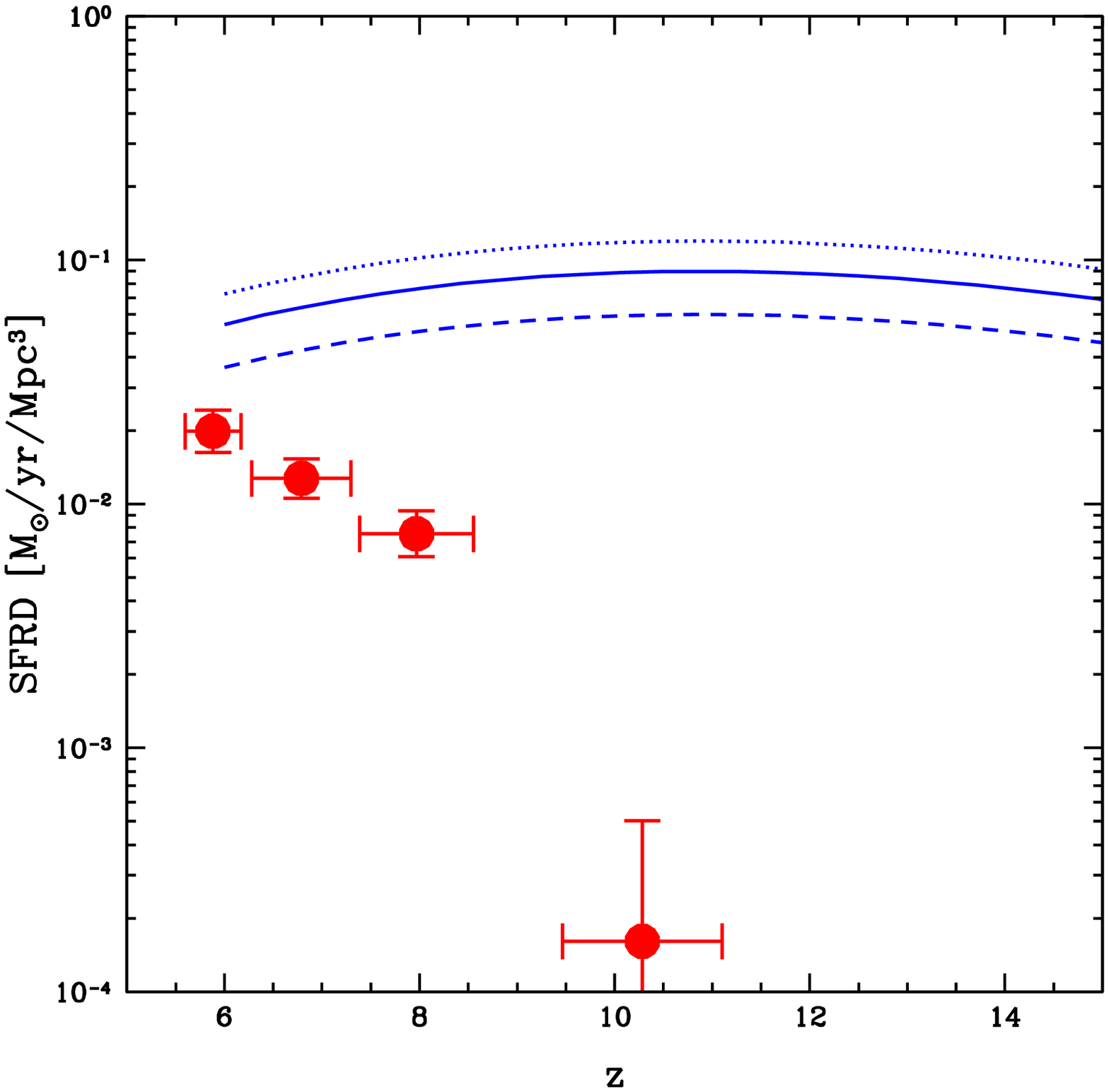}
\includegraphics[scale = 0.4]{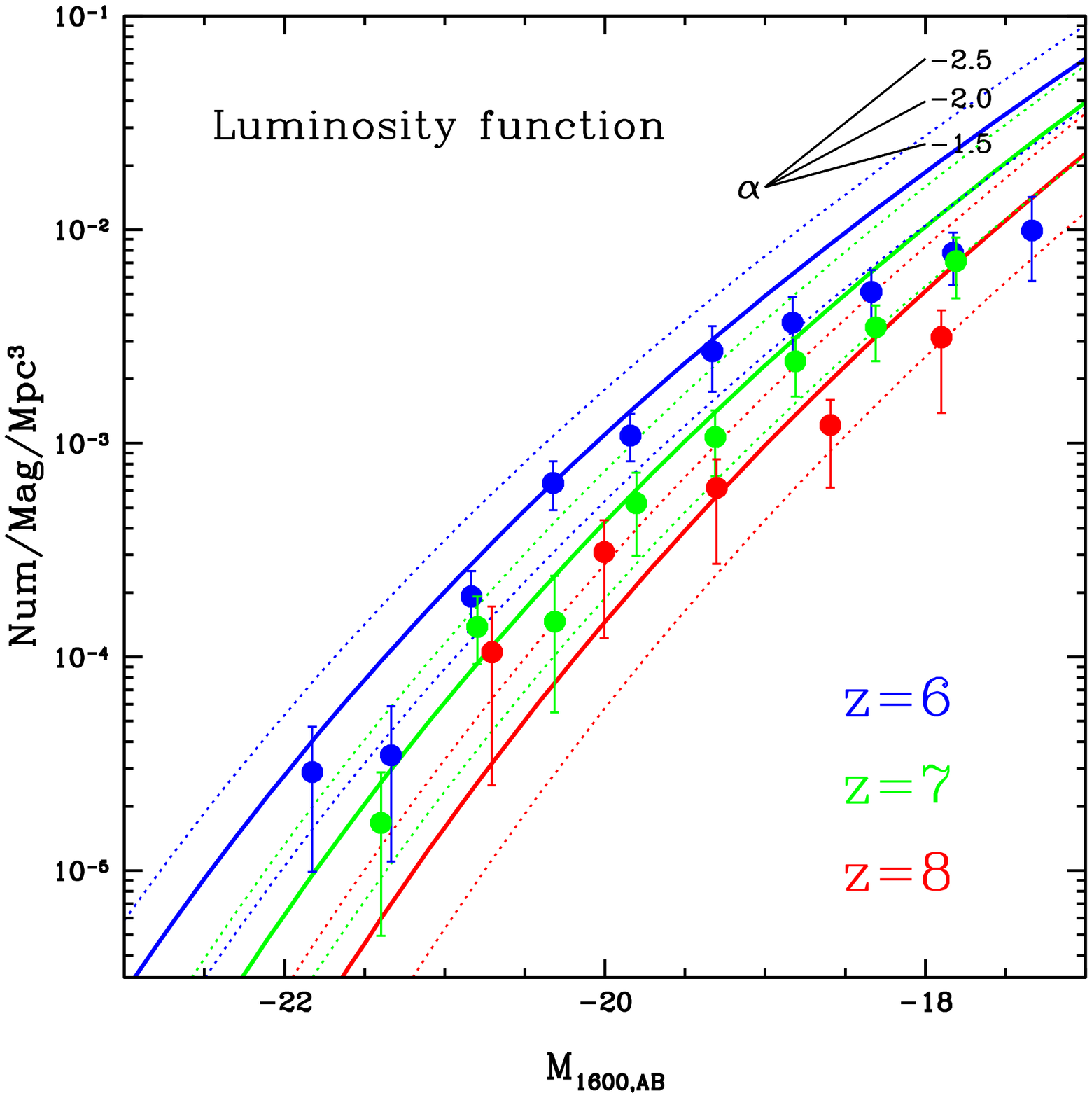}
\caption{\label{fig:LF} {\it Left:} The SFRD as a function of the redshift derived from 
Eq.(\ref{eq:psi}) for three reionization histories $\tau=0.077$ (blue dashed line), 0.090 
(blue solid line) and 0.099 (blue dotted line) which are obtained by setting $f_*=0.02$,
0.03 and 0.04, respectively. The red points are the measurements from HUDF09+ERS+CANDELS observations 
\citep{Bouwens11b}. {\it Right:} The estimated UV luminosity function at $z=6$, 7 and 8 with 
$\lambda_{\rm res}=1600\ \rm \AA$. The central thick solid lines are derived from our model with
$f_*=0.03$ at $z=6$, 7 and 8, and the thin dotted lines are for $f_*=0.04$ (upper) and 0.02
(lower) which can match the $1\sigma$ error of the data given in \citep{Bouwens11a}. 
The faint-end slope $\alpha$ at three values 
are also shown in black solid lines for comparison.
}
\end{figure*}

In order to relate our galaxy population responsible for reionization to the observations, 
we also compare our model to observations of $z > 6$ galaxies, focusing on the UV LFs. 
To obtain the UV LF, instead of the occupation number which is luminosity independent, we make use of the conditional luminosity function (CLF) approach
\citep{Yang03, CoorayM05, Cooray05}. To compute the CLF we map galaxy rest-frame UV luminosity to halo mass 
 with some scatter added  similar to the case of low-redshift galaxy populations (Cooray 2005).
The total luminosity of a halo is taken to be
\be
L_{\rm tot}(M,z) = l_{\nu}(z)f_*\frac{\Omega_{\rm b}}{\Omega_{\rm M}}M \, ,
\ee
and we assume that this total luminosity can be ascribed to the central galaxy when $M_{\rm min} < M < M_{\rm sat}$, following the'
earlier description related to the occupation number. When $M > M_{\rm sat}$ we introduce satellites with central galaxy luminosity kept fixed at $L_{\rm tot}(M_{\rm sat},z)$.
However, when comparing to the existing measurements we found that all of the rest-UV LF measurements are in the range where central galaxies dominate the LF and thus our comparison
to the measured LFs is independent of assumptions related to the exact form of the satellite occupation number or conditional luminosity function.

To compare with existing rest-UV LF measurements, we  convert the luminosity of each galaxy to the AB absolute magnitude via the relation 
$M_{\rm AB} = -2.5{\rm log_{10}}L_{\nu}+5.48$ \footnote{http://www.ucolick.org/~cnaw/sun.html}.
In the left panel of Fig.~\ref{fig:LF}, we show the SFRD as a function of redshift derived from 
Eq.(\ref{eq:psi}) for three reionization histories with $\tau=0.077$, 0.090 and 0.099 which are obtained by 
setting $f_*=0.02$, 0.03 and 0.04, respectively. The red points are the data from HUDF09+ERS+CANDELS observations 
\citep{Bouwens11b}. We find the SFRD of the three cases are higher than the existing measurements, especially at the high redshifts.
This difference is mainly due to the fact that existing SFRD estimates are limited to galaxies with $M_{\rm UV} < -17$, while the
bulk of the reionization UV density budget is contained in the galaxies at the faint-end of the LF. This is especially the case at high redshift since
the faint-end slope of the LF is steep with values reaching close to -2 already.

In the right panel of Fig.~\ref{fig:LF}, we show our rest-UV LF (corresponding to $\lambda_{\rm res}=1600\ \rm \AA$) 
at $z=$6, 7 and 8.  The central thick solid curves are the luminosity function derived from our default model with $f_*=0.03$ and 
$\tau=0.090$, and the thin dotted lines are obtained by $f_*=0.02$ (lower) and 0.04 (upper) which could match the
$1\sigma$ errors of the data given in \citep{Bouwens11a}. The three values of the faint-end slope $\alpha=-2.5$, 
-2.0 and -1.5 are shown, which indicates the slope of our model is between -2.5 and -2.0. Also, we find the 
star formation time scale $t_{\rm SF}$ is around $6\times 10^8$ yrs at $z=6$, 7 and 8 when we calculate the luminosity
mass function here.

\begin{figure}[htb]
\includegraphics[scale = 0.4]{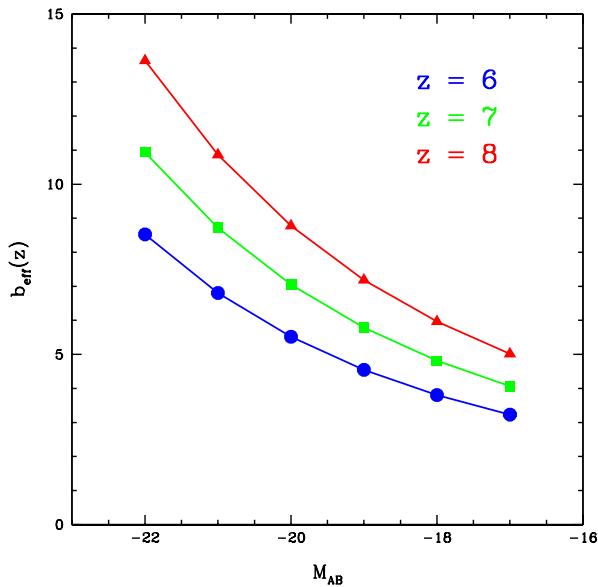}
\caption{\label{fig:b_MAB} The effective galaxy bias vs. AB magnitude $M_{\rm AB}$ at $z=6$, 7
and 8. We find the bias increases at higher redshifts and lower $M_{\rm AB}$.
}
\end{figure}

We also explore the dependence of the galaxy bias on the $M_{\rm AB}$ in Fig.~\ref{fig:b_MAB}. We can define a 
effective galaxy bias here from the HOD model as
\be
b_{\rm eff}(z) = \int_{M_{\rm min}}^{M_{\rm max}} dM \frac{dn}{dM} b(M,z)\frac{\langle N_{\rm gal} \rangle}{\bar{n}_{\rm gal}}\ ,.
\ee
In the plot we show the $b_{\rm eff}$ as a function of $M_{\rm AB}$ at $z=$6, 7 and 8 for the 
$M_{\rm AB}=-22\sim -17$. We find that the galaxy bias increases as the redshift increases and decreases as
the $M_{\rm AB}$ increases. The reason is obvious that the galaxy number density $\bar{n}_{\rm gal}$ defined
by Eq.~(\ref{eq:ngal}) becomes smaller at higher redshift and bigger at larger $M_{\rm AB}$ 
(the larger $M_{\rm AB}$ means smaller halo mass $M$).

\subsection{Anisotropy Power Spectrum}

\begin{figure*}[htb]
\includegraphics[scale = 0.4]{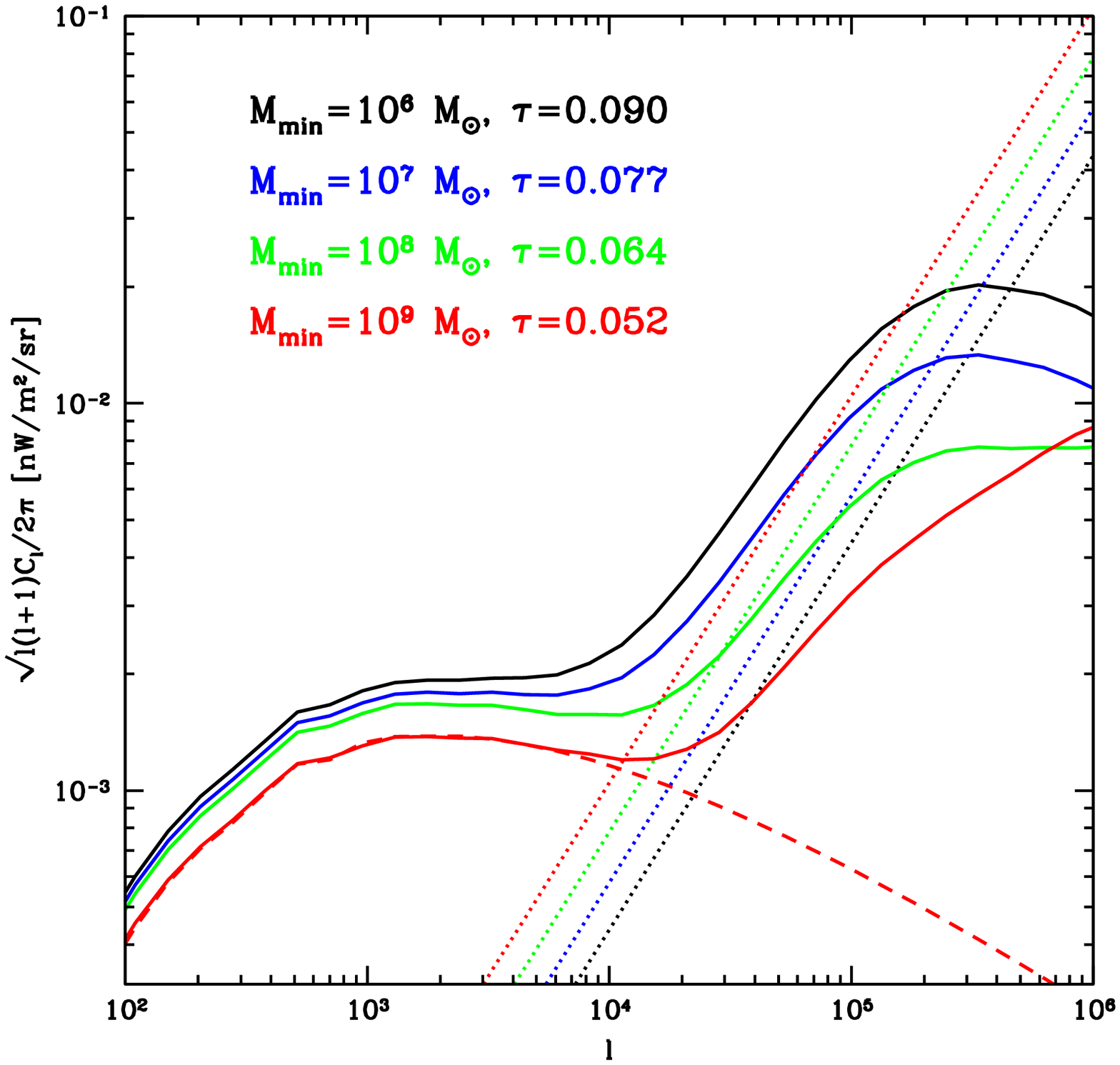}
\includegraphics[scale = 0.4]{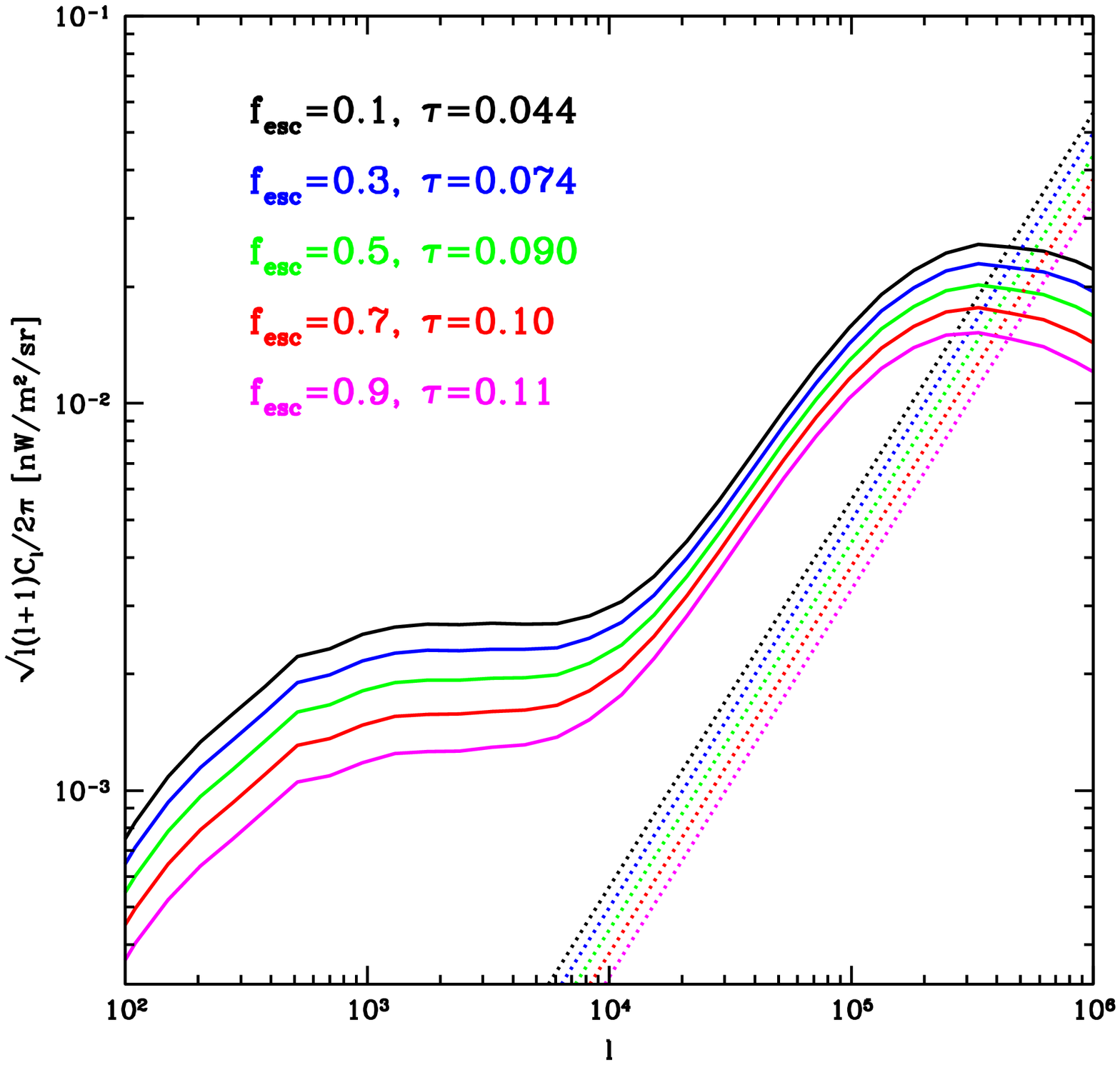}
\caption{\label{fig:Cl_Pop} The near-IR background intensity anisotropy power spectrum 
at $\lambda_{\rm obs}=3.6\ \rm \mu m$. 
The solid lines are the total clustering power spectrum with the non-linear power spectrum from the
HOD model.  {\it Left:}  We show a  variety of predictions with $M_{\rm min}$ related to the occupation
number taken to be values shown in the plot. In all of these cases we take $M_{\rm sat}=15M_{\rm min}$.
As $M_{\rm min}$ corresponds to the minimum halo mass to host a galaxy, the reionization history is also
changed and we list the optical depth to electron scattering for the four cases we have studied here.
For comparison, the linear power spectrum is  shown in a dashed line for one of the cases, which
have a turnover around $\ell=10^3$ (Cooray et al. 2004). The shot-noise power spectra are in dotted lines.
Note that the shot-noise level is higher when $M_{\min}$ is larger.
{\it Right:} Clustering predictions as a function of $f_{\rm esc}$; Note that the amplitude of the clustering power spectrum is
higher with a smaller value for $f_{\rm esc}$. However, the optical depth to electron scattering is lower with a small value of $f_{\rm esc}$.
}
\end{figure*}

In Fig.~\ref{fig:Cl_Pop} we show the near-IR background anisotropy angular power spectrum at 
$\lambda_{\rm obs}=3.6\ \rm \mu m$. The clustering power spectra
with non-linear power spectrum from the HOD model are in solid lines. The left panel shows the case with $M_{\rm min}$, the minimum mass to host a galaxy, is changed from
$10^6$ to $10^9$ M$_{\sun}$. For comparison, we also  plot the linear power spectrum which has a turnover around $\ell=10^3$. 
When calculating the clustering power spectrum, we set the parameters of the HOD with $M_{\rm min}$ values as listed in the figure
with $M_{\rm sat}=15M_{\rm min}$. The corresponding values on the optical depth to electron scattering are also listed in the figure.
We note that the shot-noise amplitude is larger for the case with  $M_{\rm min}=10^9$ M$_{\rm \sun}$ in comparison to the case with, say, $M_{\rm min}=10^6$ M$_{\rm \sun}$, though in those
two cases the clustering amplitude is higher with $M_{\rm min}=10^6$ M$_{\rm \sun}$. This is because the shot-noise amplitude is sensitive to the second flux-moment of the number counts.
By keeping the minimum mass higher we force the overall counts to be restricted to brighter sources than the case with a lower minimum halo mass. 
On the other hand the clustering power spectrum reflects the total background intensity. With the minimum mass lowered, both the overall number density
of galaxies and the background intensity are increased.

The right panel of  Fig.~\ref{fig:Cl_Pop} shows the case where we vary $f_{\rm esc}$ to highlight the fact $C_l$ amplitude is inversely proportional to $f_{\rm esc}^2$.
However, one cannot arbitrarily reduce $f_{\rm esc}$ to a small value since this results in a low optical depth to electron scattering. If $M_{\rm min} \sim 10^8$ to 10$^9$, such that
one does not need to be concerned of effects due to feedback negatively impacting the formation of galaxies in lower mass halos, then we find that $f_{\rm esc}$ must be at the level
of 0.3 or more, under the assumptions related to the IMFs for PopII and PopIII stars we use in this paper. 

\begin{figure}[htb]
\includegraphics[scale = 0.4]{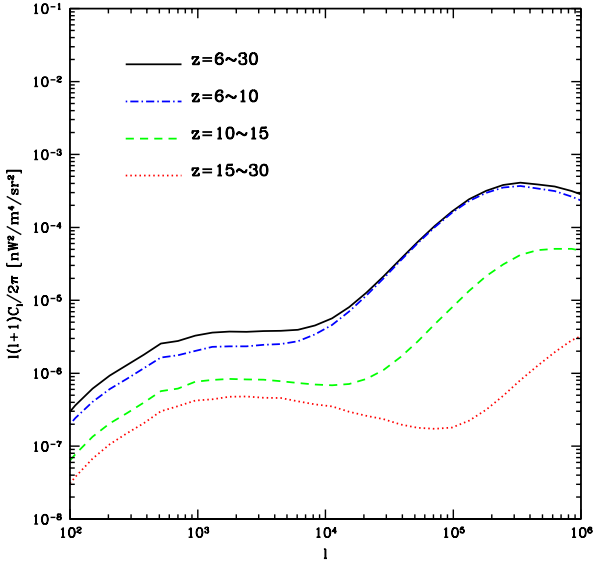}
\caption{\label{fig:Cl_zbin} The  angular power spectrum at 3.6 $\mu$m for different redshift bins.
The power spectrum dominates over the redshift range of 6 to 10. The model shown here is the default case with $\tau=0.090$ with $f_{\rm esc}=0.5$ and $f_*=0.03$.
}
\end{figure}

In Fig.~\ref{fig:Cl_zbin}, we plot the $C_{\ell}$ at different redshift bins. As can be seen, the main
contribution of the $C_{\ell}$ comes from the lowest redshift range of 6 to 10. For all practical purposes one can assume that the near-IR background is probing the end of
reionization not the first objects to form in the universe at the beginning of reionization.

\begin{figure}[htb]
\includegraphics[scale = 0.4]{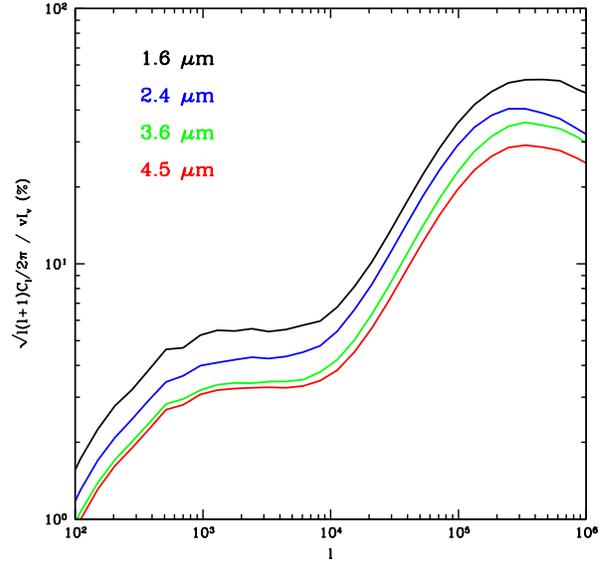}
\caption{\label{fig:Cl_Iv_ratio} The ratio of rms fluctuations to the total intensity
at different wavelengths $\lambda_{\rm obs}=1.6$, 2.4, 3.6 and
4.5 $\rm \mu m$. We find the near-IR background fluctuations are around $10 \%$ compared to the background intensity
from $\ell=10^3$ to $10^4$.
}
\end{figure}

To compare with the total near-IR intensity spectrum, we plot the ratio of the square root of $C_{\ell}$ to
the $\nu I_{\nu}$ in Fig.~\ref{fig:Cl_Iv_ratio}. 
We integrate from $z=6$ to 30 with the same HOD model described above. For easy comparison, we restrict the predictions to be
with $M_{\rm min}=10^6$ M$_{\sun}$ such that $\tau\sim0.090$ to remain consistent with WMAP 7-year result.
We find the ratio of 
$\sqrt{\ell(\ell+1)C_{\ell}/2\pi}$ to $\nu I_{\nu}$ is about $10 \%$ for $10^3<\ell<10^4$ and the clustering fluctuations amplitude is below 100\% of the intensity 
at all angular scales. This shows that large fluctuations of the background intensity  is not expected and the background behaves in a manner that is
smooth and not clumpy as in the case if spatial variations are dominated by rare, bright galaxies.

\begin{figure*}[htb]
\epsscale{1.9}
\centerline{
\resizebox{!}{!}{\includegraphics[scale = 0.37]{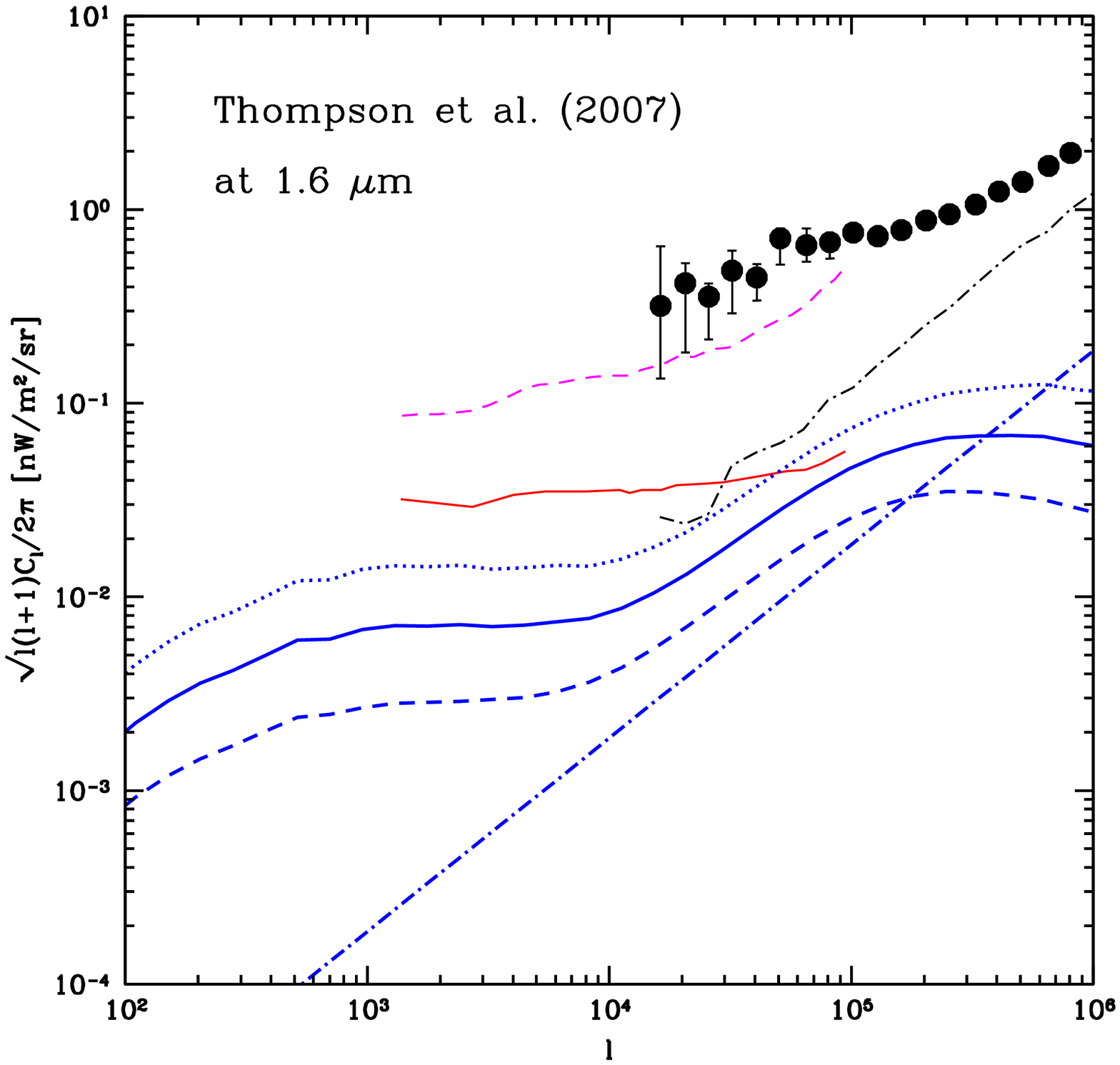}}
\resizebox{!}{!}{\includegraphics[scale = 0.37]{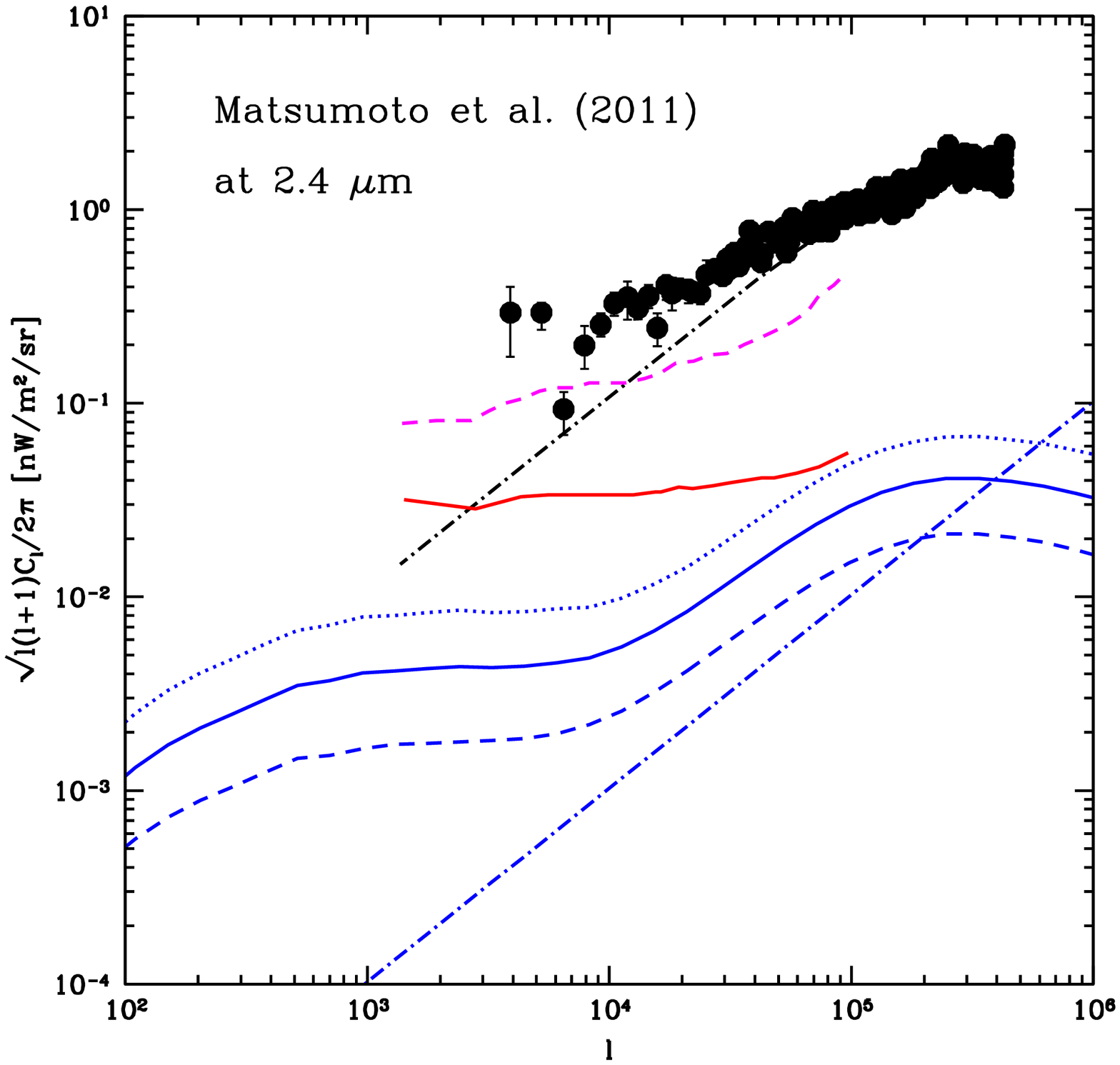}}
}
\centerline{
\resizebox{!}{!}{\includegraphics[scale = 0.37]{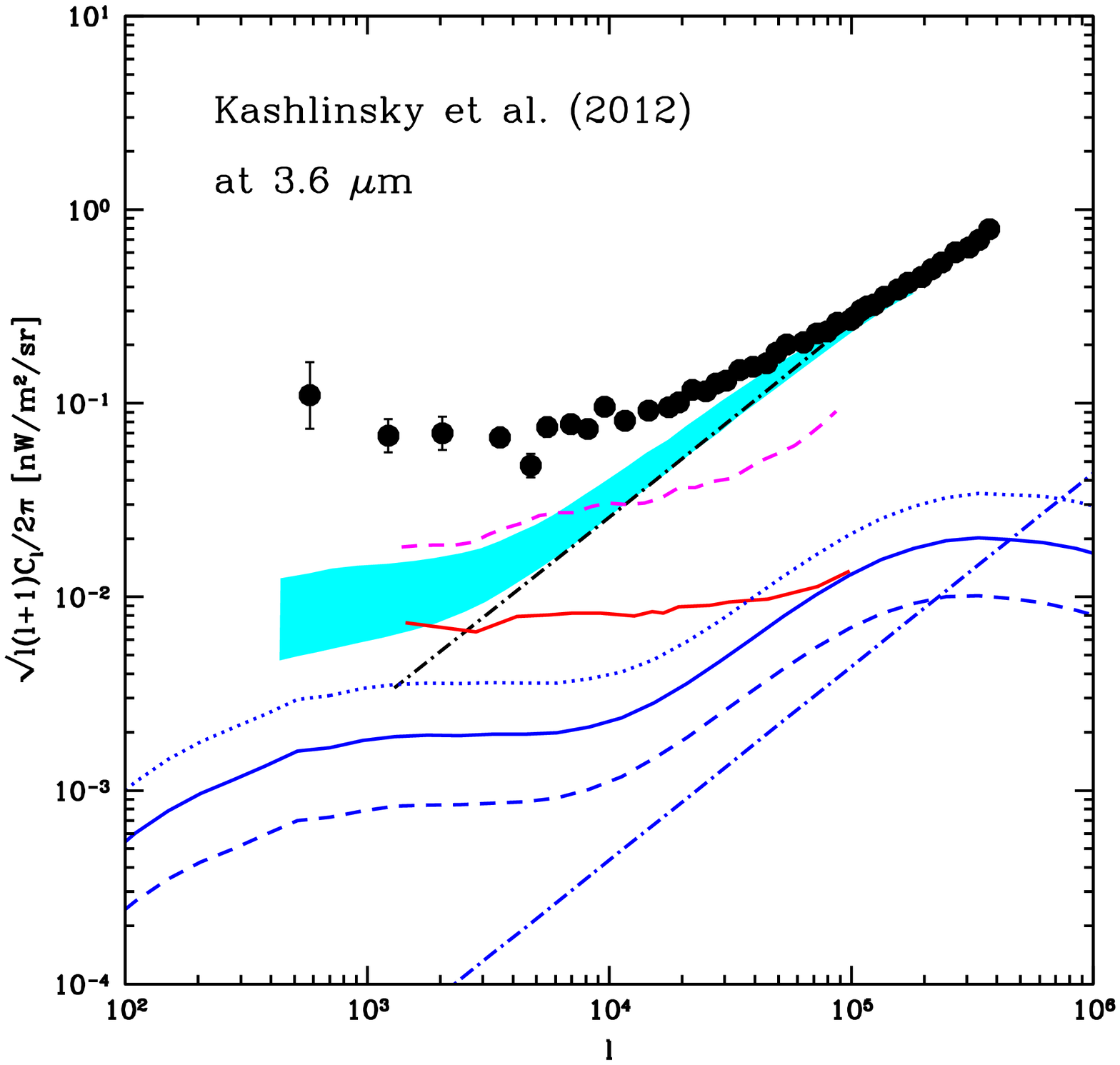}}
\resizebox{!}{!}{\includegraphics[scale = 0.37]{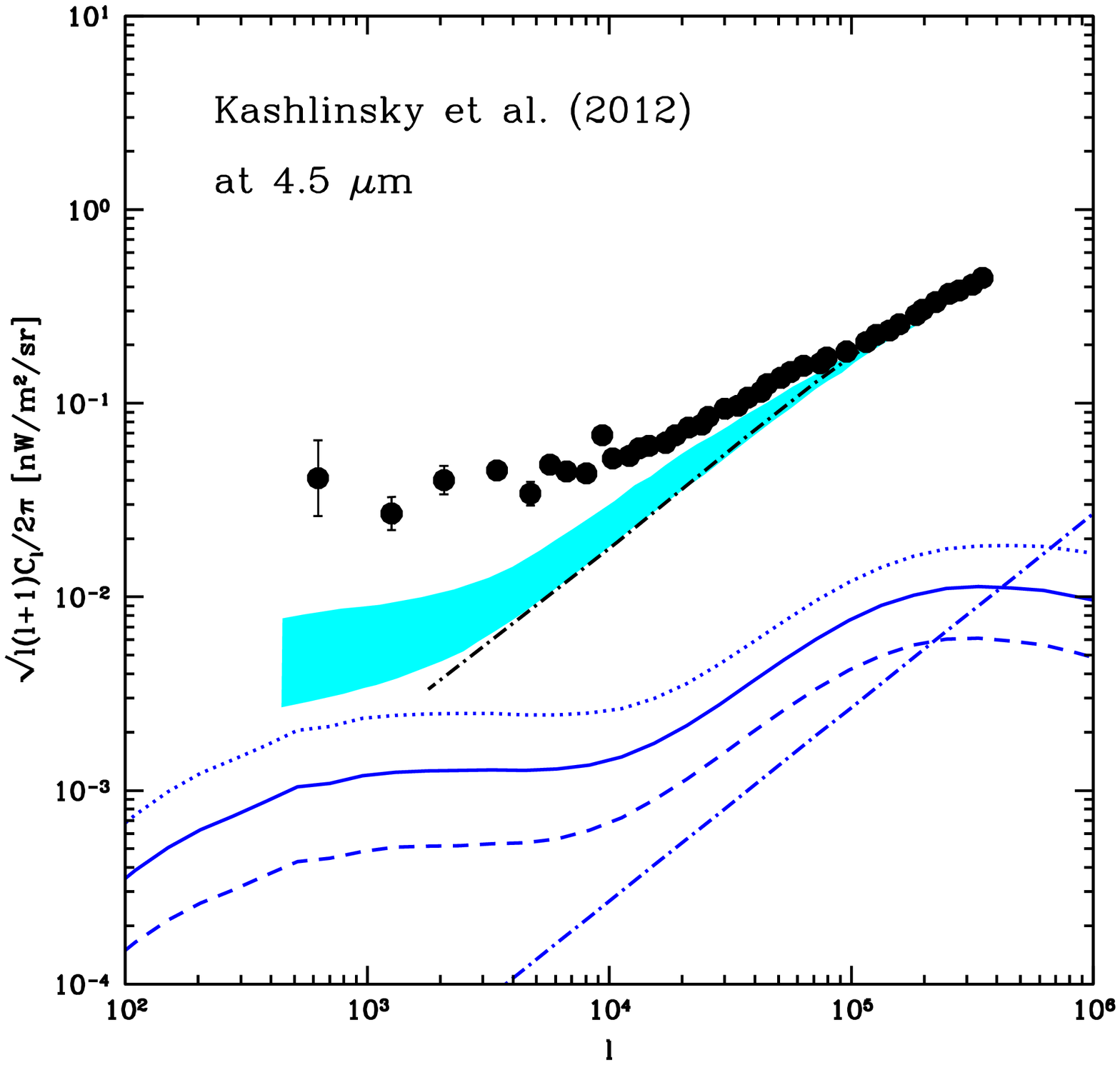}}
}
\epsscale{1.0}
\caption{\label{fig:Cl} The comparison of our $C_{\ell}$ with the current measurements at different 
wavelengths. The clustering power spectrum are in blue solid ($f_*=0.03$ and $f_{\rm esc}=0.5$), dashed 
($f_*=0.02$ and $f_{\rm esc}=0.9$) and dotted ($f_*=0.04$ and $f_{\rm esc}=0.1$) lines with $M_{\rm min}=10^6\ \rm M_{\sun}$ 
in our HOD model. The shot-noise power spectrum are in 
blue dash-dotted lines.  The observational data of shot-noise are shown in black dash-dotted lines. 
For the data of Kashlinsky et al. (2012) at 3.6 and 4.5 $\rm \mu m$, 
we removed the beam function so that a direct comparison can be made to model predictions.
The simulation results from \cite{Fernandez12} are also shown 
with minimum halo mass  set at $M_{\rm min}=10^8$ (lower red solid) and $M_{\rm min}=10^9\ \rm M_{\sun}$ (upper magenta dashed), without a suppression related to
the star-formation in  halos between the range of 10$^8$ and 10$^9$ M$_{\sun}$ where photoionization effects become important. The shaped region in each of the two lower panels
is an estimate of the residual anisotropy power spectrum signal expected from low-redshift faint galaxies following Helgason et al. (2012). These two curves clearly show the  presence of excess clustering
at $\ell \sim 10^3$. While this excess has been suggested as originating from reionization (Kashlinsky et al. 2012), we do not find this to be the case as our predictions are lower.
}
\end{figure*}

In Fig.~\ref{fig:Cl}, we show the near-IR background intensity  power spectrum (solid curves), shot-noise power 
spectrum (dotted curves) with a comparison to existing observational measurements at several wavelengths: 
$\lambda_{\rm obs}=1.6$, 2.4, 3.6, and 4.5 $\rm \mu m$ \citep{Thompson07, Matsumoto11, Kashlinsky12}.
Note that for the data from Kashlinsky et al. (2012) at 3.6 and 4.5 $\rm \mu m$, we removed the IRAC beam transfer 
function $B_{\ell}$ so the comparison to theoretical predictions is easier. An accounting of the beam is 
essential  since realistic instruments do not have perfect resolution causing a loss of power on the smallest scales. This effect
can be explicitly measured from the point spread function.  We calculate the beam transfer function $B_{\ell}$ by 
taking the power spectrum of the measured point spread function for the {\it Spitzer}/IRAC instrument, which is publicly available
and compute  $C_{\ell} \rightarrow C_{\ell}/B_{\ell}$  (for more information see Smidt et al. 2012).

We also show the results from the simulation in \cite{Fernandez12} for comparison. The red solid and magenta
dashed lines are for the simulation with  mass resolution $M_{\rm min}=10^8$ and $M_{\rm min}=10^9\ \rm M_{\sun}$, respectively. These two cases
are what is described in their work as not having a  suppression of the small mass halos (ie star-formation present in halos with mass between 10$^8$ and 10$^9$ M$_{\sun}$, where
one expects photo-ionization heating to suppress star-formation).  The reionization histories of these two cases involve PopII stars with $f_{\rm esc}=0.1$.

\begin{figure*}[htb]
\epsscale{1.9}
\centerline{
\resizebox{!}{!}{\includegraphics[scale = 0.38]{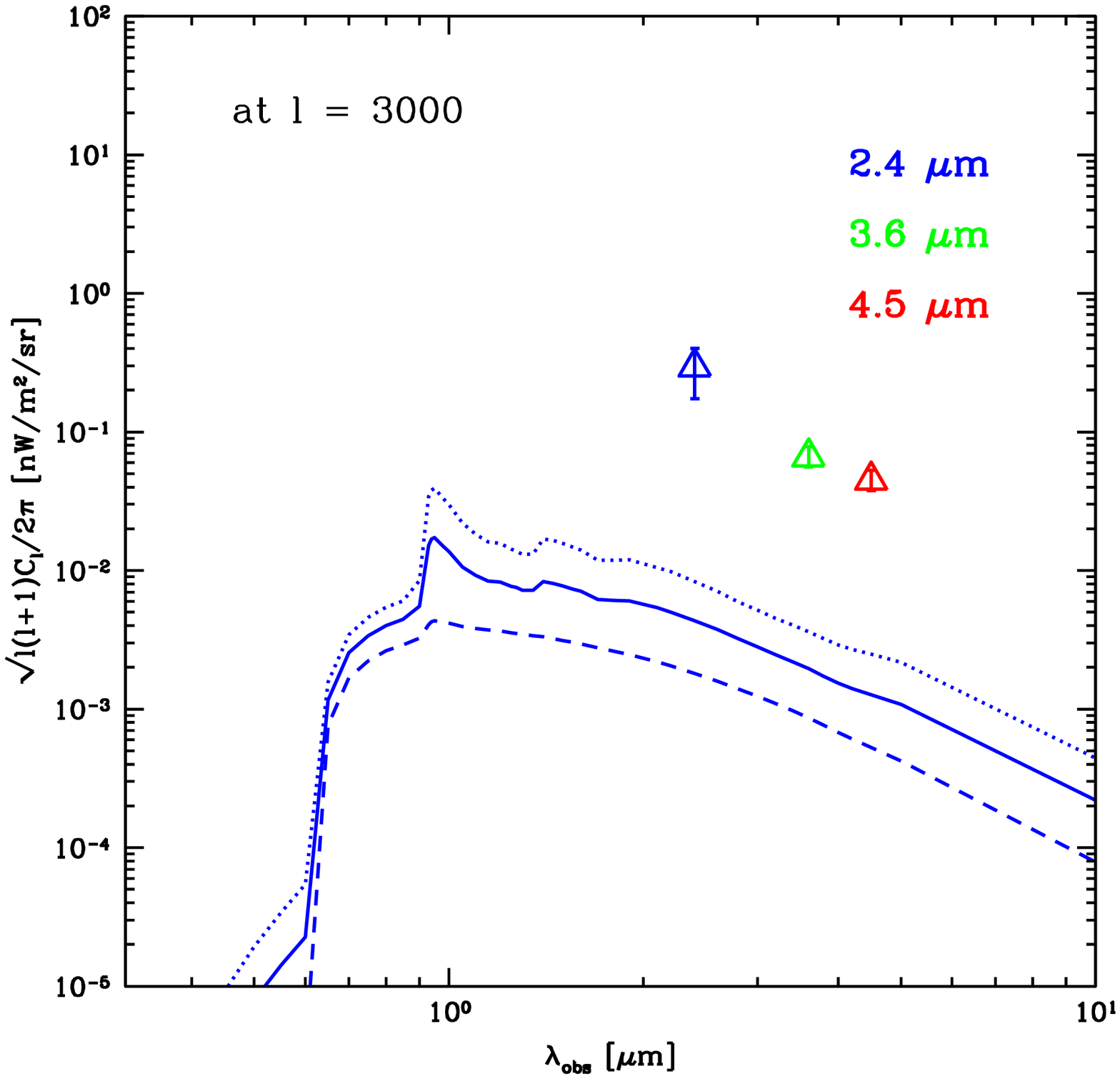}}
\resizebox{!}{!}{\includegraphics[scale = 0.38]{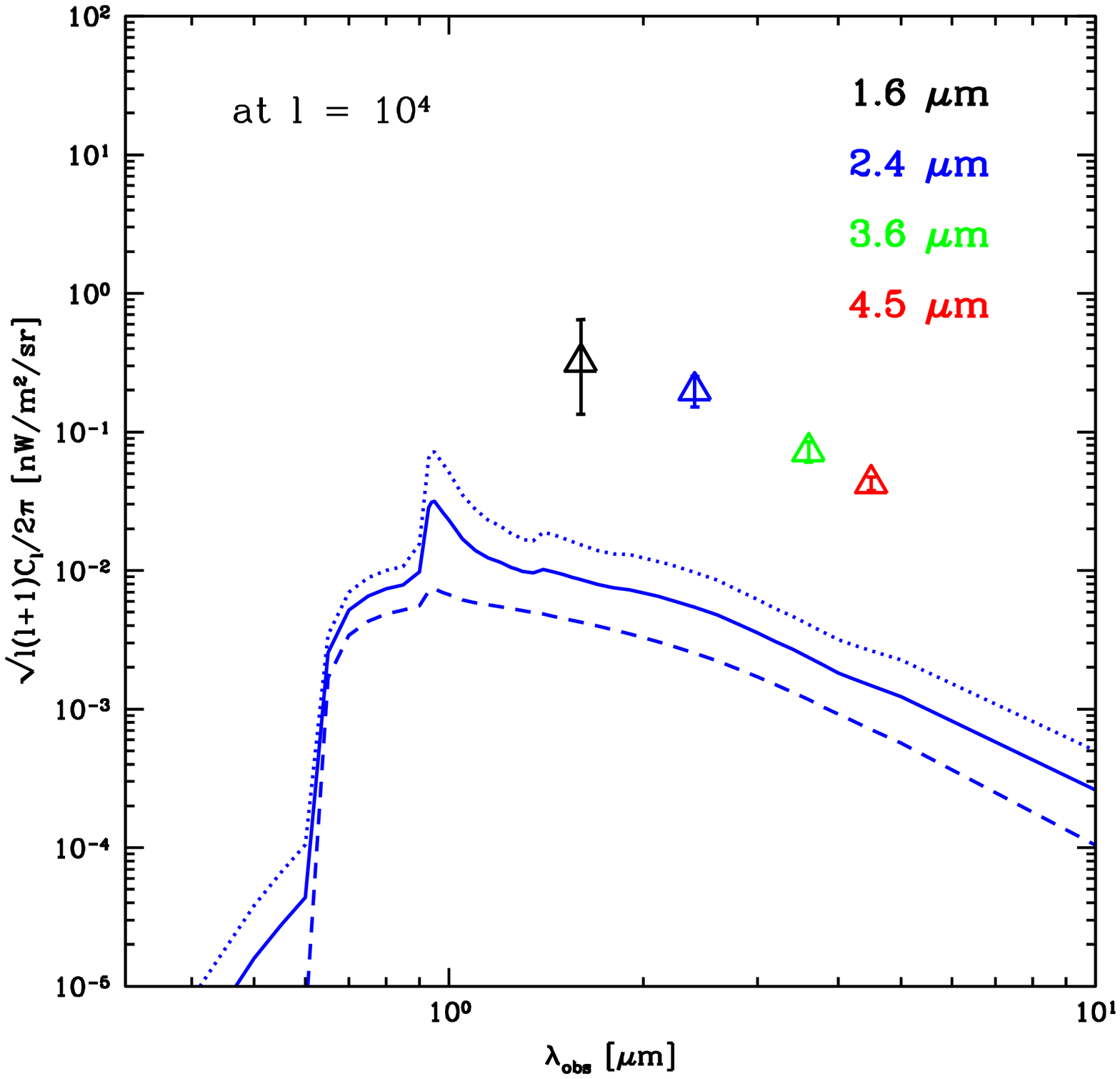}}
}
\epsscale{1.0}
\caption{\label{fig:sqtCl} The square root of the $C_{\ell}$ at $\ell=3\times 10^3$ and $\ell=10^4$ 
compared to the observational data at different wavelengths $\lambda_{\rm obs}=1.6$, 2.4, 3.6, and
4.5 $\rm \mu m$. The blue solid, dashed and dotted line are the square root of the $C_{\ell}$ with $f_*=0.03$ 
and $f_{\rm esc}=0.5$, $f_*=0.02$ and $f_{\rm esc}=0.9$, and $f_*=0.04$ and $f_{\rm esc}=0.1$, respectively.
The data points are from the data sets shown in Fig.~\ref{fig:Cl} around $\ell=3\times 10^3$ or $\ell=10^4$.}
\end{figure*}

For more clarity, we show the square root of the $C_{\ell}$ at $\ell=3\times 10^3$ and $\ell=10^4$ together with the 
data as a function of the wavelength in Fig.~\ref{fig:sqtCl}. The data points are from the
data sets shown in Fig.\ref{fig:Cl} which are around $\ell=3\times 10^3$ or $\ell=10^4$. We find the curve of the
square root of the $C_{\ell}$ has similar shape as $\nu_{\rm obs}I_{\nu}$, but the amplitude of rms fluctuations in our models
is lower than the existing measurements. We capture the uncertainties related to
$f_*$, $f_{\rm esc}$, and $M_{\rm min}$ by considering a low and high range for our prediction related to $C_l$ with $\tau$
falling within the 1 $\sigma$  uncertainty range of the WMAP 7-year result when these parameters are varied.
Even with parameter uncertainties accounted for,  we find that the existing measurements are at least an order of magnitude larger than our
model predictions. 

We attempted additional model variations but failed to find a scenario where $\tau$ is consistent with WMAP 7-year result
and the existing LFs leading to a model consistent with existing near-IR background. One can, in principle, model fit the
near-IR fluctuation power spectra by increasing the photon output of first galaxies. This results in an optical depth that is higher than the WMAP value
and a LF that has brighter galaxies than observed in existing deep HST/WFC3 images. The existing measurements require a background intensity that is
at least 3 nW m$^{-2}$ sr$^{-1}$ at 1.6 $\mu$m so that the model predictions become consistent with measurements within the 1$\sigma$
uncertainties of the measurements. Since 0.3 nW m$^{-2}$ sr$^{-1}$ at 1.6 $\mu$m  is associated with 2.5 H-ionizing photons per baryon during reionization,
if reionization is to explain the background fluctuations then we are dealing with a situation where $\sim$ 25 H-ionizing photons per baryon  are present.
A possibility is to introduce a spectrum for the emission that has a rest-frame cut-off in UV at wavelengths shortward of the Lyman-limit that is not associated with reionization but by the emission mechanism itself.
Unfortunately we have not been able to come up with such an emission spectrum.

As discussed in Helgason et al. (2012) faint, low-redshift galaxies do not have a clustering shape consistent with the
 anisotropy power spectrum measurements. The existing measurements show a clear excess in clustering at 30 arcsec to few tens arcminute angular scales
that  differ from faint galaxy clustering power spectrum. This difference is statistically significant and one can only explain at most 20\% of the near-IR fluctuations to
be associated with faint galaxies, reducing the previous estimate in Cooray et al. (2007; Chary et al. 2008) that suggested a contribution at the 50\% level.
We are now left with an unexplained set of measurements since neither the low redshift faint galaxies nor the high redshift reionization galaxies  can explain them.
One possibility is that the existing fluctuation excess in the data has a non-astrophysical origin, perhaps either involving systematics in the data or fluctuations in the Zodiacal light.
While arguments have been made that these effects are insignificant with a set of well-coordinated multi-wavelength measurements we plan to further address the origin of near-IR fluctuations in our upcoming papers.

\section{Summary}

The UV emission from stars which are formed in the early Universe from $z=6$ to 30 can contribute to the near-IR background light.
By measuring the intensity and anisotropies of the near-IR background, we can investigate the properties of these early stars
and the epoch of reionization. In this work we discuss several sources that contribute to the near-IR background intensity, including the emission of
stars, Ly$\alpha$, free-free, free-bound and two-photon. We first estimate the frequency spectrum separately for Pop II and Pop III
star with the redshift range $z=6\sim 30$. By using the initial stellar mass spectrum and the fitting model of time-averaged 
hydrogen photoionization rate we calculate the luminosity mass density. We find that although the luminosity mass density of the 
Pop III stars is a bit larger than that of Pop II stars, the near-IR intensity spectrum from the Pop II stars is stronger than that from Pop III stars, 
which is caused by the longer lifetime of Pop II stars.

In order to check the consistence of our stellar model and the reionization history, we derive the hydrogen reionization fraction at
different redshifts and calculate the optical depth by assuming the Pop II stars are dominant for $z\lesssim10$ while Pop III 
stars for $z\gtrsim 10$. We find that if we set the star formation efficiency $f_*=0.03$, the Universe would be totally reionized 
around $z=9$ with the optical depth $\tau=0.090$, which is well consistent with the result from the WMAP 7-year data. Also, we 
explore the other possible models with $f_*=0.02$ and $0.04$ and get $\tau=0.077$ and 0.099, respectively. The total number of
the ionizing photon per baryon required to maintain the ionized IGM, $N_{\rm ion}^p(z)$, is also evaluated for the three cases, 
and we find the $N_{\rm ion}^p(z)$ becomes a constant with a value around 2.5 after $z=15$.

To compare with existing bright-end luminosity function measurements, we evaluate the UV luminosity function from our 
model at $\lambda_{\rm res}=1600\ \rm \AA$ for $z=$6, 7 and 8 from $M_{\rm AB}=-22$ to $-17$. We compare to the measurements 
of Bouwens et al. (2012). We find our derived luminosity function is consistent with the data with the slope of the faint-end 
$\alpha\sim -2$. This is a steep slope and existing measurements do suggest that the slope is steeper than for LFs at low redshifts. 
We then define an effective galaxy bias $b_{\rm eff}$ from the HOD model, and find the $b_{\rm eff}$ becomes
bigger at high redshift and bright luminosities because of the lower number density of the galaxies. 

Finally, we calculate the  angular power spectrum of the near-IR background by making use of a halo model. The non-linearities provided by the 1-halo term increase
the clustering strength at multipoles greater than about $\ell \sim 2\times 10^3$. The suggest turn-over at this scale with the linear power spectrum alone in Cooray et al. (2004)
is no longer present. We find the shot-noise power spectrum of the Pop III stars is greater than that of Pop II stars because of the larger luminosity mass density for Pop III stars. 
By making use of the stellar population evolution model, we calculate the near-IR anisotropy power spectra $C_{\ell}^{\nu}$ at different wavelengths
and compare to the observational data. We find our results are lower than the observational data by at least an order of magnitude. There are now strong arguments that the
near-IR anisotropies cannot originate from low redshift faint galaxies (Helgason et al. 2012).  We have failed to explain the alternative origin of near-IR background anisotropies  involving
 galaxies present during reionization, contrary to suggestions in the literature (Kashlinsky et al. 2012). 
In future works, using additional measurements with {\it Spitzer} and {\it Hubble}/WFC3, we plan to further discuss  the near-IR fluctuations and to explain if the origin is astrophysical or
whether it is associated with yet-unknown systematic effect in the data.

\begin{acknowledgments}
This work was supported by NSF CAREER AST-0645427 and NASA NNX10AD42G at UCI. We thank CIBER, SDWFS, and CANDELS teams
for helpful discussions and questions that motivated this paper.
\end{acknowledgments}

\end{document}